\documentstyle[preprint,eqsecnum,aps]{revtex}

\begin{document}
\draft

\preprint{HD--TVP--97/03}

\title{A Numerical Study of an Expanding Plasma of Quarks in a Chiral Model}

\author{P.~Rehberg and J.~H\"ufner}
\address{Institut f\"ur Theoretische Physik, Universit\"at Heidelberg, \\
         Philosophenweg 19, D--69120 Heidelberg, Germany}

\maketitle
\begin{abstract}
We numerically solve the transport equations for a quark gas described
by the the Nambu--Jona-Lasinio model. The mean field equations of
motion, which consist of the Vlasov equation for the density and the
gap equation for the mean field, are discussed, and energy and momentum
conservation are proven. Numerical solutions of the partial
differential equations are obtained by applying finite difference
methods. For an expanding fireball of the light quark mass evolves from
small values initially to the value of 350~MeV. This leads to a
depletion of the high energy part of the quark spectrum and an
enhancement at low momenta. When collisions are included one obtains
an equation of the Boltzmann type, where the transition amplitudes
depend on the properties of the medium.  These equations are given for
flavor $SU(3)$, i.\,e. including strangeness. They are solved
numerically in the relaxation time approximation and the time evolution
of various observables is given.  Medium effects in the relaxtion times
do not significantly influence the shape of the spectra. The mass of
the strange quark changes little during the expansion. The strangeness
yield and the slope temperatures of the final spectra are studied as a
function of the size of the initial fireball.
\end{abstract}


\section{Introduction} \label{introsec}
Deconfinement and chiral symmetry determine the properties of strongly
interacting matter at high temperatures and densities to a large
extent.  While deconfinement is yet far from being understood, chiral
symmetry restoration and the resulting effects can be shown to play the
dominant role for the low energy sector of strong interactions and can
be well studied within phenomenological models.  However, most of these
studies are restricted to the case of thermal equilibrium. Since heavy
ion experiments are non-equilibrium phenomena, there is a
need for incorporating chiral symmetry breaking and restoration into a
non-equilibrium theory. This is the purpose of this paper.

In order to develop a theoretical model for the evolution of heavy ion
collisions including a dynamical breaking of chiral symmetry, one has
to choose a model Lagrangian for the underlying interaction. Here we
employ the Nambu--Jona-Lasinio (NJL) model in its two and three flavor
versions \cite{sandi}. Although nonconfining and nonrenormalizable, this
model describes well the properties of light hadrons in the vacuum.
Recent lattice results indicate that also at high temperatures around
the phase transition the temperature dependence of the condensate and
of certain masses is correctly reproduced by the model \cite{edwin}.
Its relative simplicity is an additional good reason to start from this
effective interaction.  The NJL Lagrangian is used in the
Schwinger--Keldysh formalism for non-equilibrium Green functions and
equations of motion for the particle densities are obtained. Since their
derivation is difficult, we give only a brief outline of this
calculation and refer for more details about this subject to
Refs.~\cite{zhawi,ogu}.

In the present work we focus on the numerical solution of the transport
equations. To this end we firstly investigate the mean field equations
of motion, i.\,e. the Vlasov equation, which describes the evolution of
the density evolving in a mean field, and the gap equation, which
describes the selfconsistent coupling of the mean field to the density.
We show that particle number, energy and momentum are conserved.  The
latter two conservation laws rely on the existence of a potential
energy, which we derive.  The Vlasov equation is a partial differential
equation, which we numerically solve using a finite difference
method. For simplicity, we confine ourselves to spherically symmetric
systems. The numerical solution of the Vlasov equation is presented
and the physical effects generated by the mean field are
discussed.

In a second part, binary collisions are included into the equations of
motion. Since we also want to study the generation of of strange quarks
via the processes $u\bar u\to s\bar s$ and $d\bar d\to s\bar s$, we use
the three flavor version of the NJL model. After sketching the
derivation of an equation of the Boltzmann type including medium
effects, we solve this equation using a relaxtion time approach. The
relaxation times entering this ansatz are computed from cross sections,
which depend on temperature and are given in Ref.~\cite{su3elast}. As
for the Vlasov equation, a numerical solution of the Boltzmann equation
is given.

We investigate the how the numerical results change when the medium
effects are switched off in the relaxation times and in the mean field.
We investigate also how a variation of the size of the initial system
influences the final particle spectra and multiplicities.  In
particular, we study the chemical equilibration of strange quarks as
the initial fireball grows larger.

To our knowledge, it is the first time that a kinetic equation for a
strongly interacting system is solved where the input -- mean fields
and collision terms -- are derived {\em consistently\/} from an
underlying relativistic Lagrangian and where all necessary medium
effects are included. In particular the underlying symmetry -- here
chiral symmetry -- is consistently preserved. Also in the numerical
solution we have chosen an unconventional way: Solving a partial
differential equation instead of the test particle method. Of course,
also a price has to be paid: The investigated system, a quark gas
which, at the present stage, does not have the possibility of
hadronization, is rather unrealistic and any comparison of our results
with data from heavy ion collisions contains large uncertainties.
Therefore the results of the paper are important for further
theoretical work.

\section{The Vlasov Equation for the Two Flavor NJL model}
\label{vlassec}
In this section, we give a brief derivation of the Vlasov equation and
a short introduction to the NJL model in its two flavor version and
discuss the conservation laws.  Then we outline the numerical methods
used for solving the Vlasov equation and present results for one
initial condition.

\subsection{Derivation of the Vlasov Equation}
The starting point of our discussion is the equation of motion for
the real time Green function $G^<$, which in coordinate space is
defined via \cite{KaBa}
\begin{equation}
G^<(x,y) = i \left< \bar \psi(y) \psi(x) \right> \hspace{2mm}, \label{e1}
\end{equation}
where $\psi(x)$ is is the destruction operator for a quark at
space--time point $x$, $\bar\psi(y)=\psi^\dagger(y)\gamma^0$ and
$\langle\dots\rangle$ denotes the average over the ensemble under
consideration. Note that $G^<$ is a matrix in Dirac space as well as in
color and flavor space. At the mean field level, the equation of motion for
$G^<(x,y)$ can be written as \cite{KaBa,wojtek}
\begin{equation}
\left(i {\partial \!\!\! /}_x - m_q(x) \right) G^<(x,y) = 0 \hspace{2mm}.
\label{e2}
\end{equation}
In this equation, $m_q(x)$ is a space--time dependent mass, which
is given as the sum of the bare mass and the Hartree part
of the self energy \cite{ogu,KaBa}. We transform Eq.~(\ref{e1}) to
phase space with the help of the Wigner transformation
\begin{equation}
G^<(x,p) = \int d^4u \, e^{ipu}
G^<\left(x + \frac{u}{2}, x - \frac{u}{2} \right)
\end{equation}
and introduce the quasiparticle approximation for $G^<(x,p)$
\cite{ogu,KaBa}:
\begin{eqnarray}
G^<(x,p) &=& \frac{i\pi}{2E(x,\vec p)}
\frac{\delta_{ff'}\delta_{cc'}}{N_cN_f} \left( p \!\!\! / + m_q(x)\right)
\Big[ \delta(p_0-E(x,\vec p)) n_q(x, \vec p) \label{quasi} \\ \nonumber
& & \hspace{2cm} - \delta(p_0+E(x,\vec p))
\left(2N_cN_f-n_{\bar q}(x, - \vec p) \right) \Big]
\hspace{2mm},
\end{eqnarray}
where $E(x,\vec p) = \sqrt{\vec p^2 + m_q^2(x)}$. In Eq.~(\ref{quasi})
color and flavor indices are shown explicitly and it is assumed that
the particle distributions for each individual degree of freedom are
independent of color, flavor and spin, so that they can be expressed by
$n_q(x, \vec p)$ and $n_{\bar q}(x, \vec p)$, which are the total
number of quarks and antiquarks per phase space cell, respectively.
Inserting Eq.~(\ref{quasi}) into Eq.~(\ref{e1}) and keeping only the
leading order in a gradient expansion leads to the Vlasov equation
\begin{equation}
\left[ \partial_t + \vec v (x,\vec p) \vec \partial_x
- \vec \partial_x E(x,\vec p) \vec \partial_p
\right] n_q(x, \vec p) = 0 \label{vlasov} \hspace{2mm},
\end{equation}
which is valid for both $n_q$ and $n_{\bar q}$. Here the velocity is
defined by
\begin{equation}
\vec v(x, \vec p) = \frac{\vec p}{E(x,\vec p)} = \vec \partial_p E(x,\vec p)
\label{velodef} \hspace{2mm}.
\end{equation}
Note that Eq.~(\ref{vlasov}) has been derived without making
assumptions about the interaction. It is thus generic for any model
\cite{KaBa}.

\subsection{The Mass of the Constituent Quark in the NJL Model}
In order to be complete, Eq.~(\ref{vlasov}) has to be supplemented by
an equation for the quark mass $m_q(x)$, which in turn has to be given
by some model for the interaction. In the following we will use the
$SU_f(2)$ version of the NJL model, which is defined by the Lagrangian
\cite{sandi}
\begin{equation}
{\cal L} = \bar\psi\left(i\partial \!\!\!/ - m_{0q}\right) \psi
+ G \left[ \left(\bar\psi\psi\right)^2 +
\left(\bar\psi i\gamma_5\vec\tau\psi\right)^2 \right] \hspace{2mm},
\end{equation}
where $m_{0q}$ is the current quark mass, $G$ a coupling constant and
$\vec\tau$ the Pauli matrices in flavor space. For the calculation of
the space--time dependent mass $m_q(x)$, we confine ourselves to a
selfconsistent Hartree approximation, which is the lowest order in an
$1/N_c$ expansion \cite{expand}. In this approximation, $m_q(x)$ is
given by \cite{wojtek,ogu}
\begin{equation}
m_q(x)=m_{0q} - 2iG \,\, {\rm Tr}\,iG^<(x,x) \hspace{2mm}, \label{rawgap}
\end{equation}
where the trace runs over spin, color and flavor degrees of freedom.
After inserting Eq.~(\ref{quasi}) into Eq.~(\ref{rawgap}) and
performing the $p_0$ integration, one obtains the NJL gap equation
\cite{sandi}:
\begin{equation}
m_q(x) = m_{0q} + 2G\, m_q(x) \int_{|\vec p| < \Lambda} \frac{d^3p}{(2\pi)^3}
\frac{2N_cN_f - n_q(x, \vec p) - n_{\bar q}(x, \vec p)}{E(x, \vec p)}
\hspace{2mm}. \label{gapeq}
\end{equation}
Since the NJL model is not renormalizable, an $O(3)$ cutoff $\Lambda$
has been introduced which makes the integral finite.

\subsection{Conservation Laws} \label{conserv}
The coupled equations (\ref{vlasov}) and (\ref{gapeq}) describe the
time evolution of the quark plasma in phase space. Before we describe
the numerical solution of these equations, we prove that they conserve
the total number of quarks and antiquarks
\begin{equation}
{\cal N}_q = \int \frac{d^3x \, d^3p}{(2\pi)^3} \,
n_q(x, \vec p) \qquad \qquad
{\cal N}_{\bar q} = \int \frac{d^3x \, d^3p}{(2\pi)^3} \,
n_{\bar q}(x, \vec p) \hspace{2mm}, \label{nqdef}
\end{equation}
the total momentum
\begin{equation}
\vec {\cal P} = \int \frac{d^3x \, d^3p}{(2\pi)^3} \,
\vec p \left(n_q(x, \vec p) + n_{\bar q}(x, \vec p)\right) \hspace{2mm},
\label{ptotdef}
\end{equation}
and the total energy ${\cal E}_{\rm tot}$, which will be defined later.

The time derivative of the total quark number is given, after inserting
(\ref{vlasov}) into (\ref{nqdef}), by
\begin{equation}
\frac{d{\cal N}_q}{dt} = \int \frac{d^3x \, d^3p}{(2\pi)^3}
\left( - \vec v(x, \vec p) \vec \partial_x + \vec \partial_x E(x,\vec p)
\vec \partial_p\right) n_q(x, \vec p) \hspace{2mm}. \label{dndt}
\end{equation}
After integrating the second term by parts, the two volume integrals in
Eq.~(\ref{dndt}) cancel. The remaining surface terms also vanish, if
$n_q(x,\vec p)$ vanishes both for high momenta and large values of
$|\vec x|$.  Thus the number of quarks and analogously the number of
antiquarks is conserved. Note that the conservation of the particle
numbers can be proven {\em independently\/} of the functional form of
the interaction.

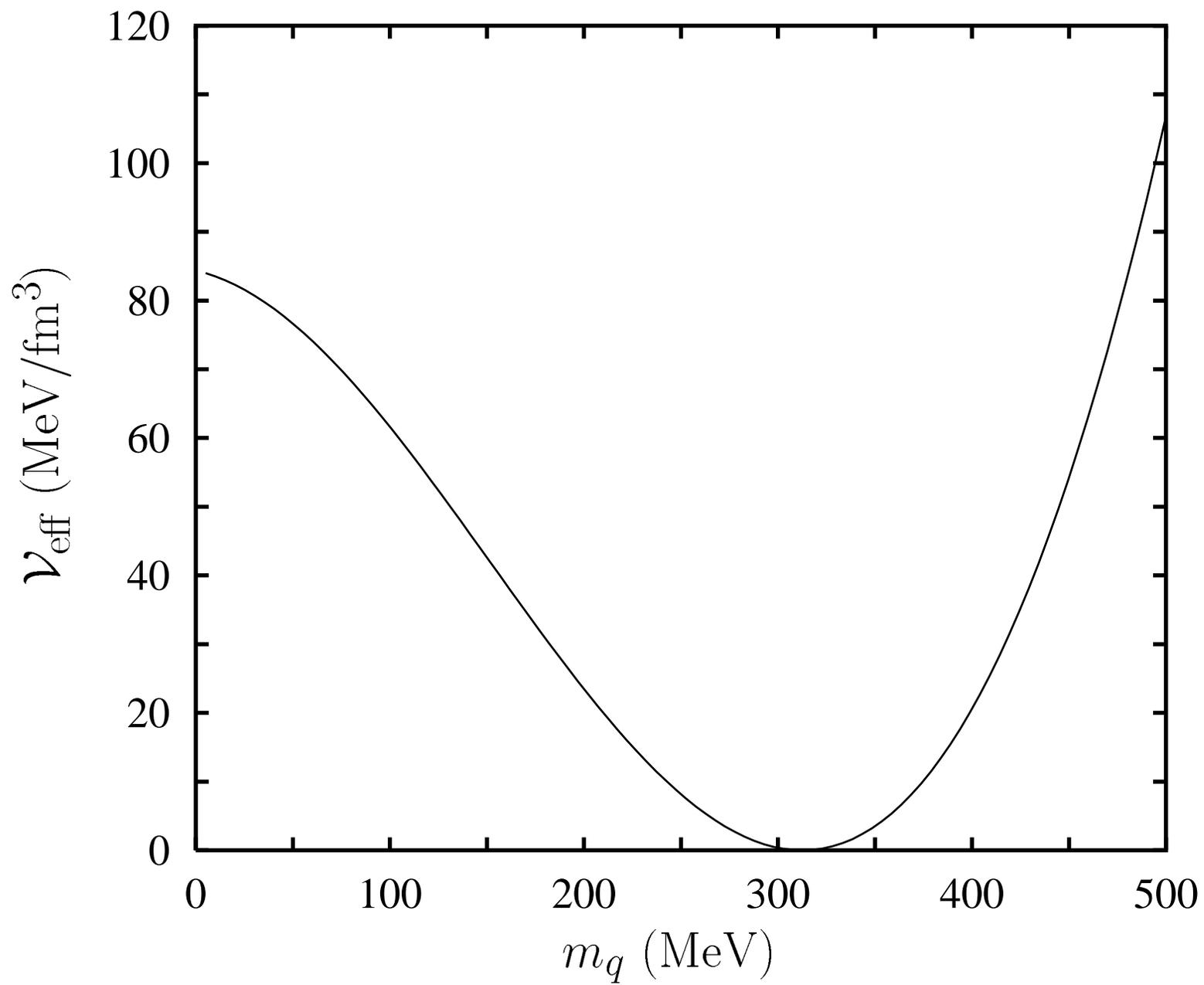
\begin{figure}
\caption[]{The effective Potential ${\cal V}_{\rm eff}$ as a function
           of the constituent quark mass $m_q$. The parameters used
           for this plot are $m_{0q}=5$~MeV, $\Lambda=653.3$~MeV and
           $G\Lambda^2=2.10$. The constituent quark mass in the vacuum
           for this parameter set is 312~MeV.}
\label{vplot}
\end{figure}

The proof of energy conservation is less trivial and depends crucially
on how $m_q(x)$ is calculated from a dynamical equation. This
is obvious since there are forms of interactions, e.\,g. explicitly
time dependent external fields, which do not conserve energy. In order
to prove energy conservation for the Vlasov equation in the NJL model,
we define the kinetic energy as
\begin{equation}
{\cal E}_{\rm kin} = \int \frac{d^3x \, d^3p}{(2\pi)^3} \,
E(x, \vec p) \left(n_q(x, \vec p) + n_{\bar q}(x, \vec p)\right)
\hspace{2mm}.
\end{equation}
After inserting the Vlasov equation (\ref{vlasov}), one obtains
\begin{eqnarray}
\frac{d{\cal E}_{\rm kin}}{dt} &=& \int \frac{d^3x \, d^3p}{(2\pi)^3}
\Big[ \partial_t E(x, \vec p)
\left(n_q(x, \vec p) + n_{\bar q}(x, \vec p)\right)
\\ \nonumber & & \hspace{2cm}
+ \left(-\vec p \vec \partial_x + m_q(x) \vec \partial_x m_q(x)
\vec\partial_p\right)
\left(n_q(x, \vec p) + n_{\bar q}(x, \vec p)\right) \Big] \hspace{2mm}.
\end{eqnarray}
The second term on the right hand side can be transformed to surface
integrals and does not contribute. The first term can be rewritten
to give
\begin{equation}
\frac{d{\cal E}_{\rm kin}}{dt} = \int d^3x \, m_q(x) \partial_t m_q(x)
\int \frac{d^3p}{(2\pi)^3} \frac{n_q(x, \vec p) + n_{\bar q}(x, \vec p)}
{E(x, \vec p)} \label{eder} \hspace{2mm}.
\end{equation}
In order to proceed further, one has to make the crucial assumption
that the system does not contain quarks with momentum larger than
$\Lambda$. Then the momentum integral in Eq.~(\ref{eder}) can be
transformed by using the gap equation (\ref{gapeq}):
\begin{equation}
\int \frac{d^3p}{(2\pi)^3} \frac{n_q(x, \vec p) + n_{\bar q}(x, \vec p)}
{E(x, \vec p)} = 2N_cN_f \int_{|\vec p| < \Lambda}
\frac{d^3p}{(2\pi)^3} \frac{1}{E(x,\vec p)} +\frac{m_{0q}-m_q}{2Gm_q}
\hspace{2mm}.
\end{equation}
The right hand side now depends on $x$ {\em only\/}
through the mass, so that we can define an effective potential ${\cal
V}_{\rm eff}$ via
\begin{equation} \label{veff}
\int \frac{d^3p}{(2\pi)^3} \frac{n_q(x, \vec p) + n_{\bar q}(x, \vec p)}
{E(x, \vec p)} = - \frac{1}{m_q(x)} \frac{d{\cal V}_{\rm eff}(m_q(x))}{dm_q}
\hspace{2mm}.
\end{equation}
The effective potential is given explicitly by \cite{su2thermo,flor}
\begin{eqnarray} \label{vexp}
{\cal V}_{\rm eff}(m_q) &=& \int^{m_q} d\mu
\left( \frac{\mu - m_{0q}}{2G} - \frac{2N_cN_f\mu}{\pi^2}
\int_0^\Lambda dp \frac{p^2}{\sqrt{p^2+\mu^2}} \right)
\\ \nonumber
&=& \frac{(m_q-m_{0q})^2}{4G} + \frac{N_cN_fm_q^4}{8\pi^2} {\rm arsinh} \!
\left(\frac{\Lambda}{m_q}\right) - \frac{N_cN_f\Lambda}{8\pi^2}
\sqrt{\Lambda^2+m_q^2}(2\Lambda^2+m_q^2) \hspace{2mm}.
\end{eqnarray}
Note that this result has been already derived in Ref.~\cite{su2thermo}
from a field theoretical point of view.  After inserting
Eq.~(\ref{veff}) into Eq.~(\ref{eder}), one obtains
\begin{equation} \label{ekons}
\frac{d{\cal E}_{\rm kin}}{dt} = - \frac{d}{dt} \int d^3x \,
{\cal V}_{\rm eff}(m_q(x))
\hspace{2mm}.
\end{equation}
By defining the potential energy ${\cal E}_{\rm pot}$ as
\begin{equation}
{\cal E}_{\rm pot} = \int d^3x {\cal V}_{\rm eff}(m_q(x))
\end{equation}
and the total energy as ${\cal E}_{\rm tot} = {\cal E}_{\rm kin} + {\cal
E}_{\rm pot}$, Eq.~(\ref{ekons}) gives
\begin{equation}
\frac{d{\cal E}_{\rm tot}}{dt} = 0 \label{ekons2} \hspace{2mm},
\end{equation}
i.\,e. the total energy is conserved. It can also be read off from
Eq.~(\ref{ekons}), that the quantity
\begin{equation} \label{epsi}
\epsilon(x) = \left[ \int \frac{d^3p}{(2\pi)^3} E(x, \vec p)
\left(n_q(x, \vec p) + n_{\bar q}(x, \vec p)\right) \right]
+ {\cal V}_{\rm eff}(m_q(x))
\end{equation}
has the meaning of an energy density. We note again, that
Eq.~(\ref{ekons2}) can only be obtained by assuming that no particles
with momentum larger than $\Lambda$ appear in the system. This somewhat
artificial restriction arises as a consequence of the introduction of
the $O(3)$ cutoff into the NJL model.  The form of ${\cal V}_{\rm eff}$
as a function of the constituent quark mass $m_q$, as evaluated from
Eq.~(\ref{vexp}), is shown in Fig.~\ref{vplot}. ${\cal V}_{\rm
eff}$ has a minimum at the constituent quark mass in the vacuum, since,
according to Eq.~(\ref{veff}), ${\cal V}_{\rm eff}$ arises as an
integral over the {\em vacuum\/} gap equation. In the vacuum, the gap
equation is thus equivalent to the condition that the energy density,
as defined by Eq.~(\ref{epsi}), is minimized. In Fig.~\ref{vplot}, we
have also added a constant to Eq.~(\ref{vexp}), so that ${\cal V}_{\rm
eff}=0$ at the vacuum quark mass and thus vanishes in the vacuum.

As for the case of energy conservation, momentum conservation can also
be proven only if the interaction is specified.  The time derivative of
the total momentum is obtained by inserting (\ref{vlasov}) into
(\ref{ptotdef}).  After performing an integration by parts and dropping
the surface terms, one obtains
\begin{equation}
\frac{d \vec {\cal P}}{dt} = - \int \frac{d^3x \, d^3p}{(2\pi)^3}
\left(n_q(x, \vec p)+n_{\bar q}(x, \vec p)\right)
\vec \partial_x E(x, \vec p) \hspace{2mm}.
\end{equation}
The structure of the momentum integral is now the same as in
Eq.~(\ref{eder}). One thus has
\begin{equation}
\frac{d \vec {\cal P}}{dt} = \int d^3x \, \vec \partial_x
{\cal V}_{\rm eff}(m_q(x)) = \vec 0 \hspace{2mm}, \label{pkons}
\end{equation}
i.\,e. the total momentum is conserved. As Eq.~(\ref{ekons2}),
Eq.~(\ref{pkons}) is only valid strictly if no particles with momentum
larger than $\Lambda$ are present.

\subsection{Numerical Methods} \label{nummet}
The Vlasov equation (\ref{vlasov}) is a first order partial
differential equation, which in general depends on the seven variables
$t$, $\vec x$ and $\vec p$. The usual solution method for this equation
is the test particle method \cite{clint,Aichelin}, which is based on a
solution of the characteristic equations. Here we do not follow this
way, but rather solve the equation by a finite difference method in
order to study the advantages and disadvantages of this approach. In
order to reduce the numerical effort, we assume that the system is
spherically symmetric. This, in turn, implies that $n_q(x, \vec p)$ can
only depend on time and the three variables
\begin{equation}
r = |\vec x| \hspace{2mm}, \hspace{2cm}
p = |\vec p| \hspace{2mm}, \hspace{2cm}
\eta = \frac{\vec x \vec p}{|\vec x||\vec p|} \hspace{2mm}.
\label{varia}
\end{equation}
The last of these variables, $\eta$, is the cosine of the angle
enclosed between $\vec x$ and $\vec p$.  With the variables of
Eq.~(\ref{varia}), Eq.~(\ref{vlasov}) transforms to
\begin{equation}
\partial_t n_q = \frac{1}{E} \left[ \eta \left(m_q \partial_r m_q \,
\partial_p n_q - p \, \partial_r n_q\right) + \left( 1 - \eta^2 \right)
\left( \frac{m_q\partial_r m_q}{p} - \frac{p}{r} \right)
\partial_\eta n_q \right] \label{varvlas} \hspace{2mm},
\end{equation}
where the arguments of $n_q$, $m_q$ and $E$ have been dropped for
simplicity. In order to solve Eq.~(\ref{varvlas}), we use a finite
difference scheme. In detail this proceeds as follows:  Firstly, the
variables $r$, $p$, $\eta$ are discretized equidistantly according to
the prescription
\begin{mathletters} \begin{eqnarray}
r_i & = & i \, r_{\rm max}/N_r  \hspace{2mm}, \hspace{3.6cm}  i = 1 \dots N_r
\\
p_j & = & j \, p_{\rm max}/N_p  \hspace{2mm}, \hspace{3.5cm}  j = 1 \dots N_p
\\
\eta_k & = & \left[ 2(k-1)/(N_\eta-1)\right] - 1 \hspace{2mm}, \hspace{1cm}
k = 1 \dots N_\eta
\hspace{2mm}.
\end{eqnarray} \end{mathletters}
This parametrization avoids the coordinate singularities of
Eq.~(\ref{varvlas}) occurring at $r=0$ and $p=0$. The time derivative
of $n_q$ at time $t_n$ is replaced by
\begin{equation}
\partial_t n_q(t,r,p,\eta) \to \frac{1}{\Delta t}\left[
n_q(t_{n+1}, r_i, p_j, \eta_k) - n_q(t_n, r_i, p_j, \eta_k) \right]
\hspace{2mm}.
\end{equation}
In order to achieve numerical stability, we use a {\em fully implicit \/}
procedure for the representation of $\partial_r n_q$ at time $t_n$,
\begin{equation} \label{discrete}
\partial_r n_q(t,r,p,\eta) \to \left\{
\begin{array}{cll}
\frac{1}{\Delta r} \left[
n_q(t_{n+1}, r_{i+1}, p_j, \eta_k) - n_q(t_{n+1}, r_{i}, p_j, \eta_k) \right]
&, \quad & i = 1 \\
\frac{1}{2\Delta r} \left[
n_q(t_{n+1}, r_{i+1}, p_j, \eta_k) - n_q(t_{n+1}, r_{i-1}, p_j, \eta_k) \right]
&, \quad & 1 < i < N_r \\
\frac{1}{\Delta r} \left[
n_q(t_{n+1}, r_{i}, p_j, \eta_k) - n_q(t_{n+1}, r_{i-1}, p_j, \eta_k) \right]
&, \quad & i = N_r
\end{array} \right.
\end{equation}
and analogous prescriptions for $\partial_p n_q$ and $\partial_\eta
n_q$.  Note that this implies open boundary conditions, i.\,e. we
assume that $n_q$ can be continued smoothly to points outside the
grid.  The time update for $n_q$ is done using an operator splitting
scheme \cite{recipes}, i.\,e. we sequentially perform the three update
steps which would arise if only the terms proportional to $\partial_r
n_q$, $\partial_p n_q$ and $\partial_\eta n_q$ were present on the
right hand side of Eq.~(\ref{varvlas}). Technically, each update
consists of the solution of tridiagonal linear systems, which can be
implemented efficiently using the Thomas algorithm \cite{recipes}. With
these ingredients at hand, the calculation proceeds as follows:
\begin{enumerate}
\item Initialize $n_q$ and $n_{\bar q}$ according to some given
      initial conditions.
\item Compute $m_q$ at each space point. \label{masscomp}
\item Compute physically relevant quantities like particle density,
      energy density etc. and store them for later evaluation.
\item Calculate $n_q$ and $n_{\bar q}$ on the next time slice.
\item Proceed with step \ref{masscomp}.
\end{enumerate}
The calculation stops at that time, when the system has become a system
of noninteracting particles. A sufficient criterion for this is that
the quark mass comes close to its vacuum value everywhere.

\subsection{Numerical Results} \label{vnumres}
In this section we discuss our numerical results for the $SU_f(2)$ Vlasov
equation. The model parameters chosen are $m_{0q} = 5.0$~MeV,
$\Lambda=653$~MeV and $G\Lambda^2=2.10$ \cite{su2elast}. The initial
conditions correspond to a spherically symmetric fireball around the
origin with baryon number zero:
\begin{equation} \label{inicon}
n_q(x, \vec p) = n_{\bar q}(x, \vec p) =
\exp\left(-\frac{r^2}{2r_0^2}\right) \,
\frac{2N_cN_f}{\exp(\sqrt{p^2+m_i^2}/T_0) + 1} \,\,
f_c\left(\frac{p+p_c - \Lambda}{\delta p}\right)
\hspace{2mm},
\end{equation}
with a cutoff function $f_c(x) = (1-\tanh x) / 2$. The first factor of
Eq.~(\ref{inicon}) describes the spatial shape of the system.  The
second factor is a Fermi distribution, which describes the distribution
in momentum space. The mass $m_i$ is chosen to be the thermal mass
corresponding to $T_0$, i.\,e. we have complete thermal equilibrium
{\em in the centre of the fireball\/}. The third factor serves to cut
off high momenta. This is necessary, since the boundary condition of
Eq.~(\ref{discrete}) assumes that the particle distribution can be
continued smoothly beyond the grid boundary. In order to avoid the
cutoff artifacts which were discussed in Section~\ref{conserv}, and
to obtain exact energy conservation, it is thus necessary to cut off
the distribution smoothly.  The function $f_c$ is designed to
interpolate smoothly between the values $f_c[(p+p_c-\Lambda)/\delta p]
\approx 1$ for $p \ll \Lambda - p_c$ and $f_c[(p+p_c-\Lambda)/\delta p]
= 0$ for $p \gg \Lambda - p_c$, the drop between these two values
taking place within an interval $\delta p$ around $p=\Lambda - p_c$.
The standard parameters chosen for Eq.~(\ref{inicon}) are $r_0=3$~fm
for the initial radius and $T_0=240$~MeV for the initial temperature.
The parameters entering the momentum cutoff factor are $p_c=100$~MeV,
$\delta p=20$~MeV.

The maximal values of $r$ and $p$ on the grid are $r_{\rm max} =
10$~fm, $p_{\rm max} = \Lambda$. The size of the grid is $N_r=N_p=100$,
$N_\eta=50$.

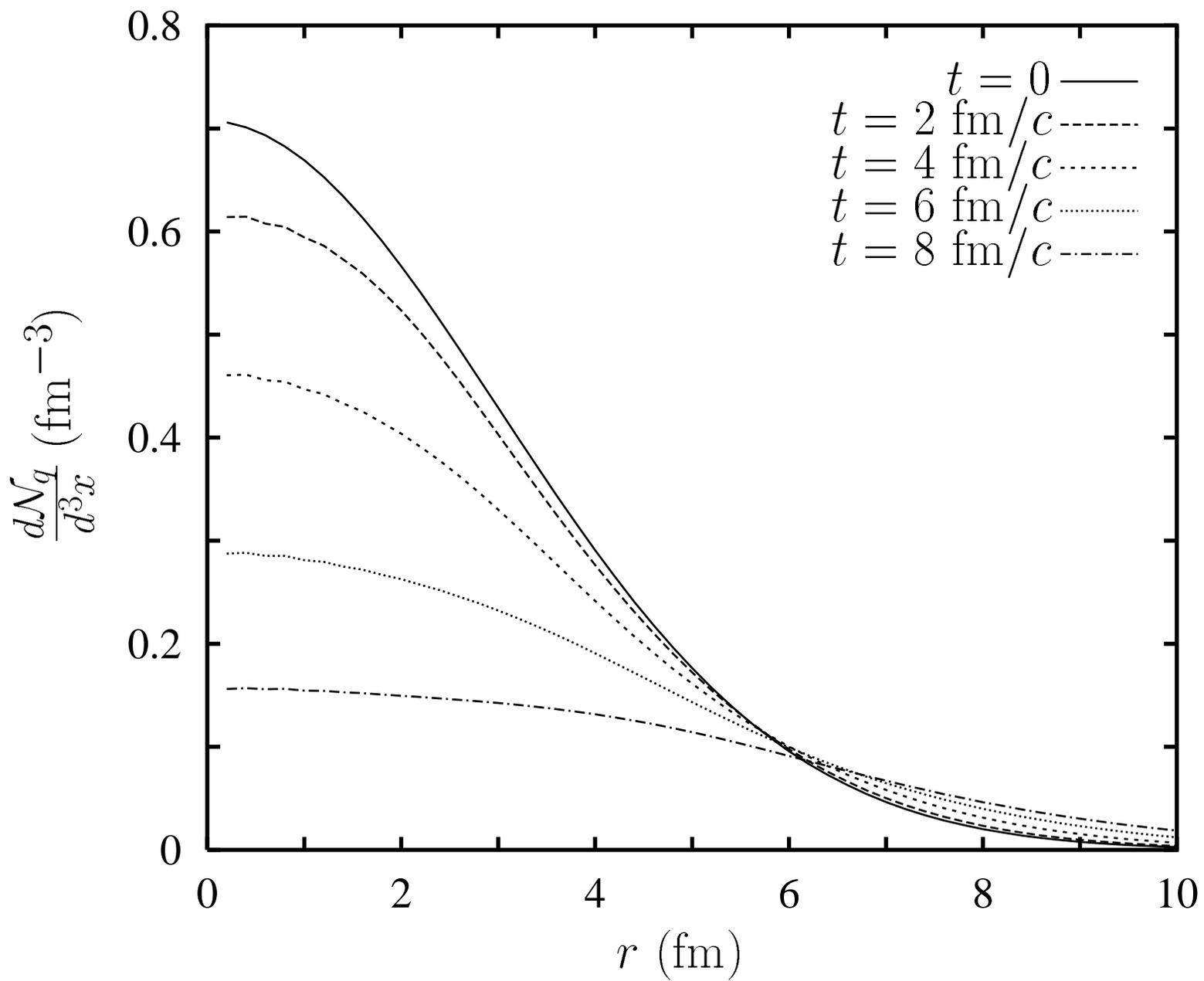
\begin{figure}
\caption[]{The radial dependence of the quarks daensity as a function
           of $r$ at various times $t$ during the expansion process.}
\label{qdens}
\end{figure}

The fireball described at $t=0$ by Eq.~(\ref{inicon}) expands for $t>0$
due to thermal motion. After a certain time the density of particles is
sufficiently low that the system behaves as a system of non interacting
particles. This is demonstrated in Fig.~\ref{qdens}, where the quark
density at various times is shown as a function of $r$. At $t=0$, the
density is described by the factor $\exp\left(-r^2/2r_0^2\right)$ in
Eq.~(\ref{inicon}). At later times, the system expands and the density
drops. Although Fig.~\ref{qdens} shows the density up to $t=8$~fm only,
the calculation has been extended to $t=17$~fm$/c$, when the density
practically vanishes.

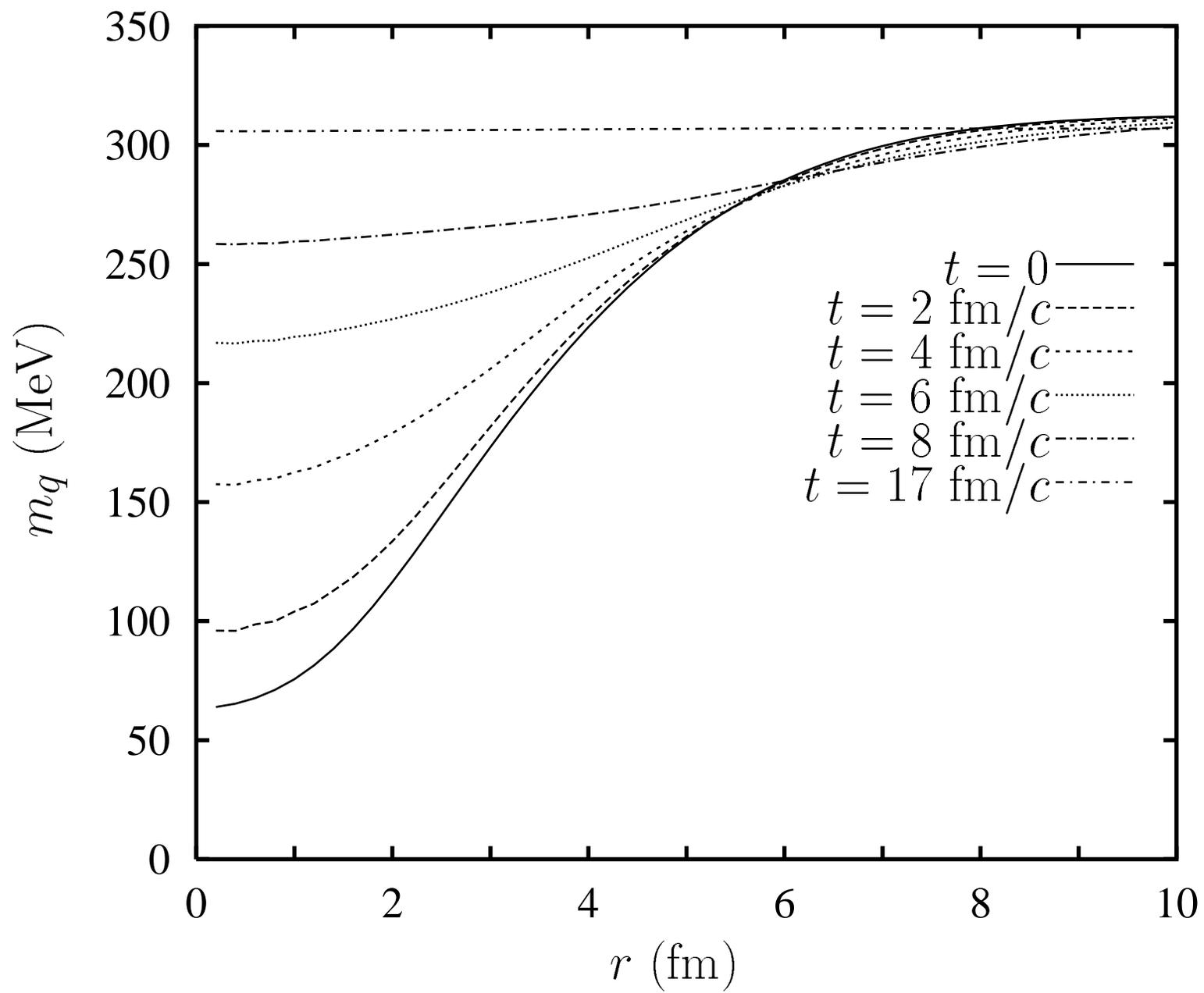
\begin{figure}
\caption[]{The radial dependence of the quark masses at various times
           $t$ during the expansion process.}
\label{qmass}
\end{figure}

The quark mass $m_q(r)$, shown in Fig.~\ref{qmass}, behaves in a
similar way. At $t=0$ the quark mass forms a ``potential well'' due to
the high density in the centre.  As the density drops, the quark mass
approaches its vacuum value of $m_{q,\rm vac}=312$~MeV. For $t\ge
17$~fm$/c$, $m_q(r)$ is practically constant. This leads to a vanishing
mean field term in Eq.~(\ref{vlasov}) and the system decouples.

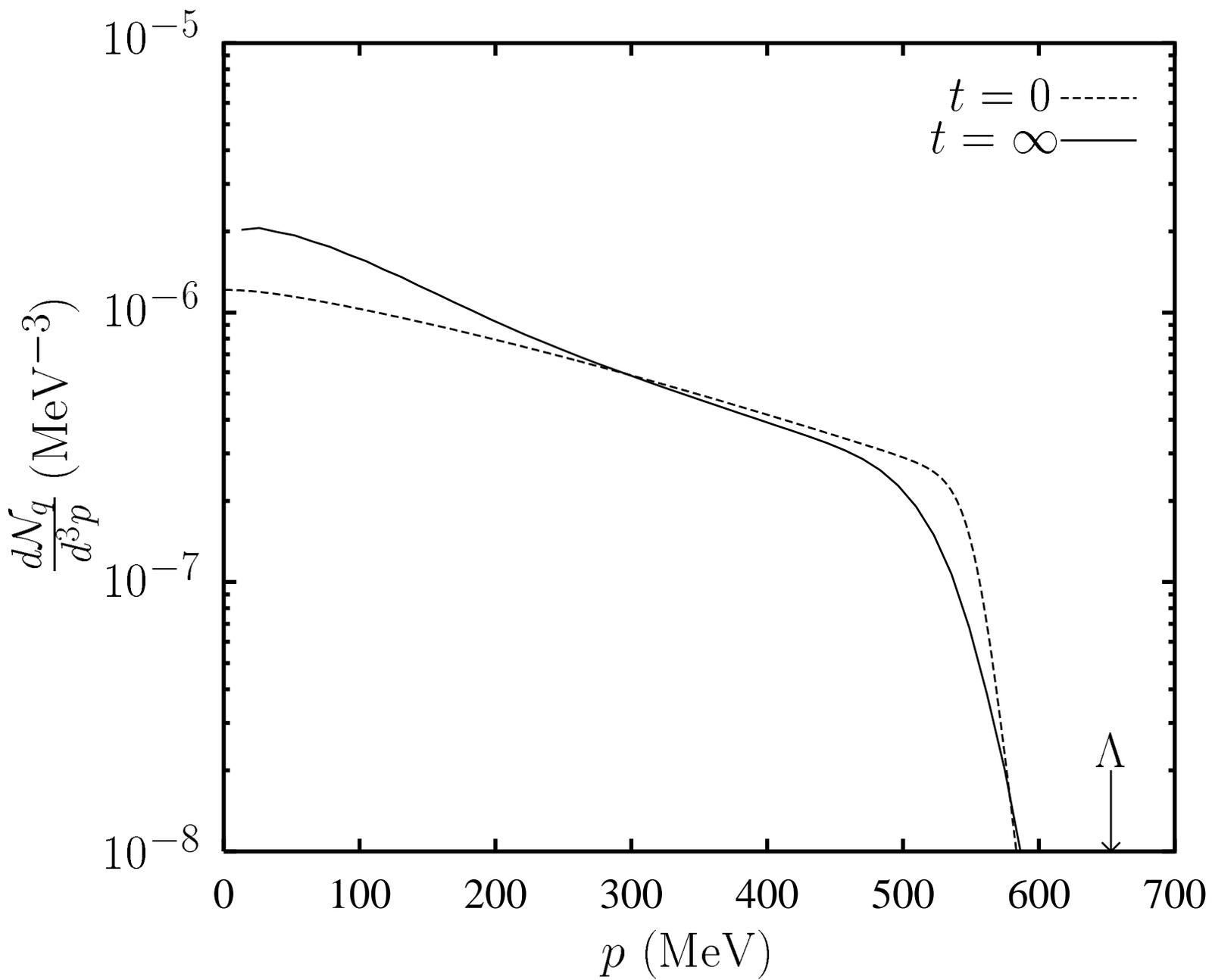
\begin{figure}
\caption[]{Momentum spectra for quarks in the beginning ($t = 0$,
	   dashed line) and at the end ($t=\infty$, solid line) of the
	   expansion.}
\label{spectra}
\end{figure}

The final ``observed'' momentum spectra are calculated using the
observation that there exist closed surfaces in coordinate space with
the properties that (i) at $t=0$ all particles are contained within
this surface, (ii) at time $t=\infty$ all particles are outside the
surface and (iii) outside the surface the system behaves effectively as
a system of non-interacting particles. The particle spectra then can be
obtained by integrating the flux through such a surface. Choosing the
surface to be a sphere with radius $r_1$, one obtains
\begin{equation}
\frac{d{\cal N}_q}{d^3p} = \frac{r_1^2}{(2\pi)^3} \int_0^\infty dt
\int_{r=r_1} d\Omega \, n_q v_\bot \hspace{2mm}, \label{flux}
\end{equation}
where $v_\bot$ is the velocity component perpendicular to the surface.
Since in practice the time integration in Eq.~(\ref{flux}) can only be
extended to finite times, we add also the (small) contribution of those
particles, which still remain inside the surface $r=r_1$ when the
calculation stops. In Fig.~\ref{spectra}, the momentum spectra of
quarks are shown for the initial and final states. At $t=0$, the
spectra are governed by the Fermi part of Eq.~(\ref{inicon}) up to
$p\approx 550$~MeV, where the distribution is cut off by the factor
$f_c\left[(p+p_c-\Lambda) / \delta p\right]$ in Eq.~(\ref{inicon}).
During the evolution, the number of particles with low momenta is
enhanced, while the high momentum region is depleted, since the
generation of a dynamical mass during the evolution slows particles
down and thus enhances the low momentum region of the spectrum.

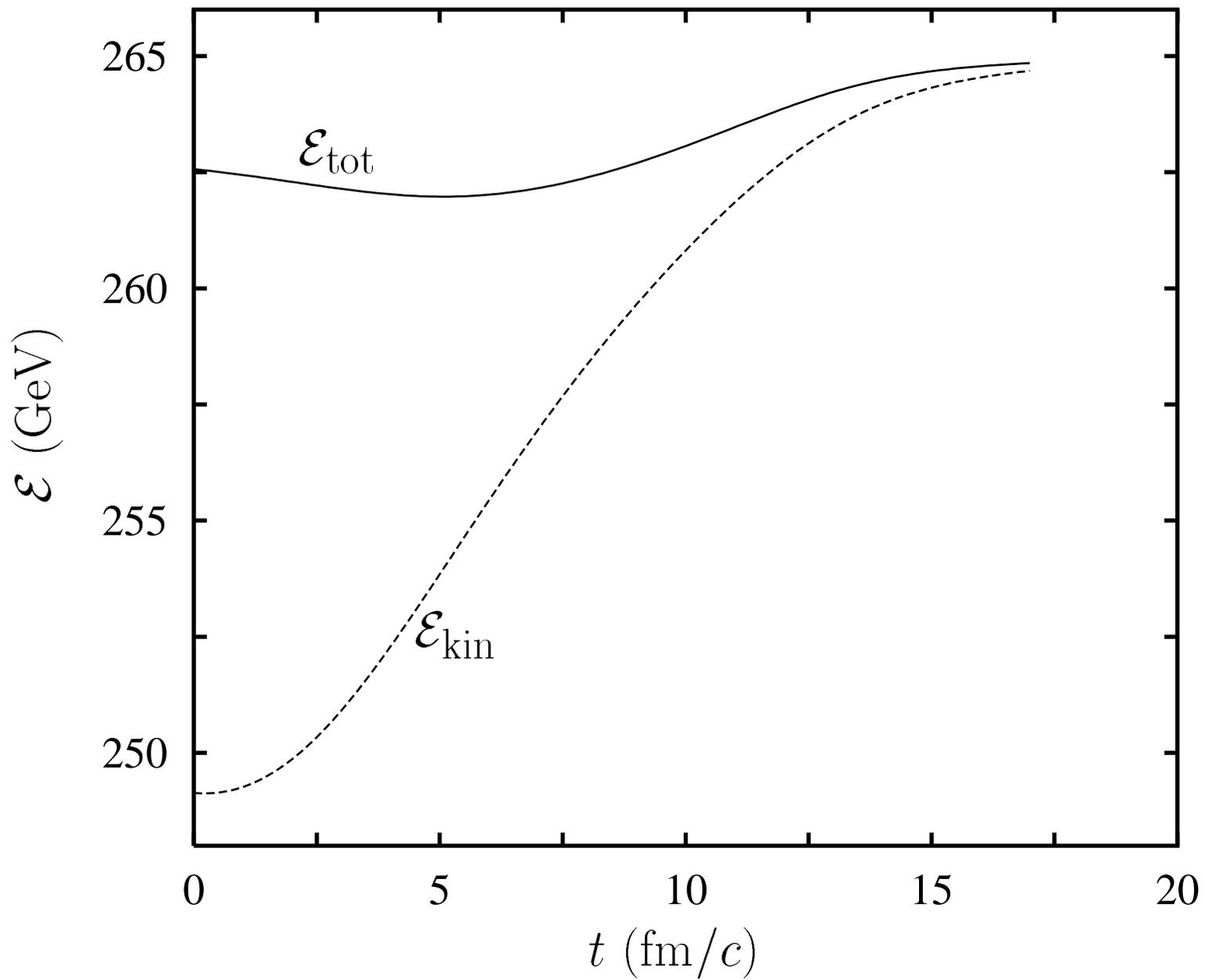
\begin{figure}
\caption[]{Time dependence of kinetic energy (dashed line) and total
           energy (solid line). Note the broken scale on the energy axis.}
\label{eplot}
\end{figure}

The time development of the total energy is interesting in two
respects: firstly as an accuracy check to the program and secondly for
a study of the importance of the potential energy, which is usually
neglected. The time behaviour of ${\cal E}_{\rm kin}$ and ${\cal
E}_{\rm tot}$ is given in Fig.~\ref{eplot}. As was detailled in Section
\ref{conserv}, the potential energy is positive and vanishes in the
vacuum. This behaviour can be observed in Fig.~\ref{eplot}: At small
times, when the system consists of low mass quarks, the potential
energy is maximal and contributes about 5\% to the total energy.
During the expansion, the quark mass approaches its vacuum value and
the the potential energy goes to zero. Concomitantly, the kinetic
energy rises. Calculating kinetic and potential energies numerically,
we find that energy is conserved to an accuracy of 1\%. We also obtain
particle number conservation with an accuracy of 6\%. Both results
point to a numerical accuracy of the order of 5--10\%. The total
momentum, on the other hand, vanishes identically due to the spherical
geometry.

\section{The Boltzmann Equation for the Three Flavor NJL Model} \label{bolsec}
The results of the previous section are generalized in two respects: by
including strange quarks and by going beyond the mean field level and
adding the collision term. The exact expression involving collision
integrals is replaced by the relaxation time approximation and
numerical results are obtained.

\subsection{Derivation}
After studying a collisionless system, we extend our approach in
order to include collisions. These can be incorporated by generalizing
Eq.~(\ref{e2}) to \cite{KaBa}
\begin{equation}
\left(i{\partial \!\!\! /}_x - m(x) \right) G^<(x,z) =
\int d^4y \, \left(\Sigma^c(x,y)G^<(y,z) - \Sigma^<(x,y)G^a(y,z) \right)
\label{stossort}
\hspace{2mm},
\end{equation}
where the causal self energy $\Sigma^c$, the anticausal Green function
$G^a$ and the self energy $\Sigma^<$ appear on the right hand side.
Since we also consider the production of strange particles, we
have to extend the underlying dynamical model. The Lagrangian for the
$SU_f(3)$ NJL model reads \cite{sandi,su3hadron}
\begin{eqnarray} \label{su3lag}
{\cal L} &=& \sum_{f=u,d,s} \bar\psi_f(i\partial\! \! \! /-m_{0f})\psi_f
       + G\sum_{a=0}^8\left[(\bar\psi\lambda^a\psi)^2 +
                          (\bar\psi i\gamma_5\lambda^a\psi)^2\right]
       \\ \nonumber &-&
        K\left[\det\bar\psi(1+\gamma_5)\psi + \det\bar\psi(1-\gamma_5)\psi
                                \right]
\hspace{2mm}.
\end{eqnarray}
Equation (\ref{su3lag}) contains two coupling constants $G$ and $K$.
The matrices $\lambda^a$ are the Gell-Mann matrices in flavor space
with $\lambda^0=\sqrt{2/3}$.
The Green function $G^<$ in the quasiparticle approximation is still
diagonal in flavor space, but no longer independent of flavor:
\begin{eqnarray}
G^<(x,p) &=& \frac{i\pi}{2E_f(x,\vec p)}
\frac{\delta_{ff'}\delta_{cc'}}{N_c} \left( p \!\!\! / + m_f(x)\right)
\Big[ \delta(p_0-E_f(x,\vec p)) n_f(x, \vec p) \label{quasit} \\ \nonumber
& & \hspace{2cm} - \delta(p_0+E_f(x,\vec p))
\left(2N_c-n_{\bar f}(x, - \vec p) \right) \Big]
\hspace{2mm},
\end{eqnarray}
where $n_f(x, \vec p)$ is the density of quarks of flavor $f$\/ for
$f=u$, $d$, $s$ and $E_f(x, \vec p)= \sqrt{\vec p^2+m_f^2(x)}$.  The
mean field derived from the Lagrangian (\ref{su3lag}) becomes
\cite{sandi,su3hadron}
\begin{eqnarray} \label{su3gap}
m_f(x) = m_{0f} + 4 G &m_f(x)& \int_{|\vec p| < \Lambda}
\frac{d^3p}{(2\pi)^3}
\frac{2N_c-n_f(x, \vec p)-n_{\bar f}(x, \vec p)}{E_f(x, \vec p)}
\\ \nonumber + 2K &m_{f'}(x)&
\int_{|\vec p| < \Lambda} \frac{d^3p}{(2\pi)^3}
\frac{2N_c-n_{f'}(x, \vec p)-n_{\bar f'}(x, \vec p)}{E_{f'}(x, \vec p)}
\\ \nonumber \times
&m_{f''}(x)&\int_{|\vec p| < \Lambda} \frac{d^3p}{(2\pi)^3}
\frac{2N_c-n_{f''}(x, \vec p)-n_{\bar f''}(x, \vec p)}{E_{f''}(x, \vec p)}
\hspace{2mm},
\end{eqnarray}
where $f$, $f'$ and $f''$ are pairwise distinct flavors.

After performing a Wigner transformation and a gradient expansion of
Eq.~(\ref{stossort}), inserting Eq.~(\ref{quasit}) and taking the real
part of the resulting equation, one arrives at a Boltzmann like equation
\cite{ogu}
\begin{equation} \label{boltz}
\left(\partial_t + \vec v_f \vec \partial_x
- \vec \partial_x E_f \vec \partial_p \right) n_f(x, \vec p) =
\sum_{f_1, f', f_1'} \left(I^{\rm coll}_{ff_1\to f'f_1'}
+ I^{\rm coll}_{f\bar f_1\to f'\bar f_1'} \right) \hspace{2mm},
\end{equation}
where $I^{\rm coll}_{ff_1\to f'f_1'}$ and $I^{\rm coll}_{f\bar f_1\to
f'\bar f_1'}$ are the collision integrals due to quark--quark
scattering $ff_1\to f'f_1'$ and quark--antiquark scattering $f\bar
f_1\to f'\bar f_1'$, respectively. One has
\begin{eqnarray} \label{qqloss}
& & I^{\rm coll}_{ff_1\to f'f_1'} =
\frac{1}{2E_f} \int dQ
(2\pi)^4 \delta^4(p + p_1 - p' - p_1')
\frac{1}{2} \overline{|{\cal M}|^2}_{ff_1\to f'f_1'}
\\ \nonumber & & \hspace{7mm}
\left[ n_{f'}(x, \vec p') n_{f_1'}(x, \vec p_1')
\phi_f(x, \vec p) \phi_{f_1}(x, \vec p_1) -
n_f(x, \vec p) n_{f_1}(x, \vec p_1)
\phi_{f'}(x, \vec p') \phi_{f_1'}(x, \vec p_1') \right] \hspace{2mm},
\end{eqnarray}
with the statistical factor $1/2$ appearing in front of the squared
transition amplitude. In Eq.~(\ref{qqloss}) we have used the
invariant integration volume
\begin{equation}
dQ = \frac{d^3p_1}{(2\pi)^3 2E_{f_1}} \frac{d^3p'}{(2\pi)^3 2E_{f'}}
\frac{d^3p'_1}{(2\pi)^3 2E_{f_1'}}
\end{equation}
and the blocking factors
\begin{equation}
\phi_f(x, \vec p) = 1 - \frac{n_f(x, \vec p)}{2N_c} \hspace{2mm}.
\end{equation}
In contrast to Eq.~(\ref{qqloss}), the collision integral due to
quark--antiquark scattering does not contain the statistical factor.
One has
\begin{eqnarray}
& & I^{\rm coll}_{f\bar f_1\to f'\bar f_1'} =
\frac{1}{2E_f} \int dQ
(2\pi)^4 \delta^4(p + p_1 - p' - p_1')
\overline{|{\cal M}|^2}_{f\bar f_1\to f'\bar f_1'}
\label{qqbloss} \\ \nonumber & & \hspace{7mm}
\left[ n_{f'}(x, \vec p') n_{\bar f_1'}(x, \vec p_1')
\phi_f(x, \vec p) \phi_{\bar f_1}(x, \vec p_1) -
n_f(x, \vec p) n_{\bar f_1}(x, \vec p_1)
\phi_{f'}(x, \vec p') \phi_{\bar f_1'}(x, \vec p_1') \right] \hspace{2mm}.
\end{eqnarray}
Note that, in terms of the differential cross section, one has
\cite{BD}
\begin{equation}
\frac{1}{2E_f} dQ (2\pi)^4 \delta^4(p + p_1 - p' - p_1')
\kappa \overline{|{\cal M}|^2}
= \frac{d^3p_1}{(2\pi)^3} d\Omega \, v_{\rm rel} \frac{d\sigma}{d\Omega}
\hspace{2mm},
\end{equation}
where $\kappa$ is a statistical factor which equals $1/2$ for the
elastic scattering of identical particles and 1 for all other processes.
Equations (\ref{boltz}), (\ref{qqloss}) and (\ref{qqbloss}) form the
basis of our following investigations.

\subsection{Relaxation Time Approximation}
The collision integrals in Eq.~(\ref{boltz}) contain the differential
cross sections $d\sigma/d\Omega$ for elastic scattering processes like
$uu\to uu$ and inelastic ones like $u\bar u\to s\bar s$. The scattering
amplitude ${\cal M}$ in Eqs.~(\ref{qqloss}), (\ref{qqbloss}) depends on
the selfconsistent masses $m_f(x)$ and on the actual densities $n_f(x,
\vec p)$ via the Pauli factors \cite{su3elast,su3hadron}.  Thus ${\cal
M}$ has to be recalculated at every space--time point of the expansion.
This is a formidable numerical task and goes beyond the scope of this
paper. Therefore we use the relaxation time approximation. Our ansatz
for the transport equation is
\begin{mathletters} \label{relaxa} \begin{eqnarray}
\left(\partial_t + \vec v_q \vec\partial_x - \vec\partial_x E_q
\vec\partial_p \right) n_q(x, \vec p) &=& \label{qrelaxa}
\frac{\tilde n_q(x, \vec p) - n_q(x, \vec p)}{\tau_{qq}(x, \vec p)}
- \frac{n_q(x, \vec p)}{\tau_{qs}(x, \vec p)} + G_q(x, \vec p) \\
\left(\partial_t + \vec v_s \vec\partial_x - \vec\partial_x E_s
\vec\partial_p \right) n_s(x, \vec p) &=& \label{srelaxa}
\frac{\tilde n_s(x, \vec p) - n_s(x, \vec p)}{\tau_{ss}(x, \vec p)}
- \frac{n_s(x, \vec p)}{\tau_{sq}(x, \vec p)} + G_s(x, \vec p) \hspace{2mm}.
\end{eqnarray} \end{mathletters}
As in Section~\ref{vlassec}, we set $n_u(x, \vec p) = n_d(x, \vec p) =
n_q(x, \vec p) / 2$ and also $n_f(x, \vec p) = n_{\bar f}(x, \vec p)$.
The first term on the right hand side of Eq.~(\ref{qrelaxa}) describes
gain and loss terms due to the elastic scattering of light quarks. In
this term, $\tilde n_q(x, \vec p)$ denotes an effective equilibrium
distribution function to be detailled below.  The second term describes
the loss of light quarks due to the processes $u\bar u\to s\bar s$ and
$d\bar d\to s\bar s$. The last term models the gain of light quarks
from strange quarks due to the reverse processes. Its concrete form
will also be given below. The terms on the right hand side of
Eq.~(\ref{srelaxa}) have similar origins, with light and strange quarks
being suitably exchanged.  The relaxation times $\tau_{qq}$,
$\tau_{ss}$, $\tau_{qs}$ and $\tau_{sq}$ are calculated from the
equilibrium cross sections \cite{su3elast}. They will be given in
Section \ref{relsec}.

The effective densities $\tilde n_q(x, \vec p)$ and $\tilde n_s(x, \vec
p)$ are chosen to have the form
\begin{mathletters} \label{neff} \begin{eqnarray}
\tilde n_q(x, \vec p) &=&
\frac{4N_c}{\exp\left(p_\mu u_q^\mu(x) / T(x)\right) + 1} \\
\tilde n_s(x, \vec p) &=& \label{zeta_und_mordio}
\zeta(x)\frac{2N_c}{\exp\left(p_\mu u_s^\mu(x) / T(x)\right) + 1} \hspace{2mm},
\end{eqnarray} \end{mathletters}
where $u_f^\mu(x)$ is the collective velocity of the fluid component
with flavor $f$ at space--time point $x$, $T(x)$ an effective flavor
independent temperature and $\zeta(x)$ a factor, which measures the
amount of chemical equilibrium for the strange quarks. The velocity
$u_f^\mu(x)$ can be calculated from the numerical solution $n_f(x, \vec
p)$ at every time slice,
\begin{equation}
u_f^\mu(x) = {\int \frac{d^3p}{(2\pi)^3} \frac{p^\mu}{E} n_f(x, \vec p)}
\Biggm/ {\int \frac{d^3p}{(2\pi)^3} n_f(x, \vec p)} \hspace{2mm},
\end{equation}
so that $T(x)$ and $\zeta(x)$ remain to be determined.

The requirement of energy conservation is used to fix the
effective temperature $T(x)$ and the density factor $\zeta(x)$.  After
introducing a potential energy which, as for the $SU_f(2)$ case,
compensates the change of the kinetic energy due to the mean field, one
obtains
\begin{eqnarray} \label{bekons}
\frac{1}{2} \frac{d{\cal E}_{\rm tot}}{dt} &=&
\int \frac{d^3x\,d^3p}{(2\pi)^3} \frac{E_q(x, \vec p)}{\tau_{qq}(x, \vec p)}
\left(\tilde n_q(x, \vec p) - n_q(x, \vec p) \right)
\\ \nonumber &+&
\int \frac{d^3x\,d^3p}{(2\pi)^3} \frac{E_s(x, \vec p)}{\tau_{ss}(x, \vec p)}
\left(\tilde n_s(x, \vec p) - n_s(x, \vec p) \right)
\\ \nonumber &+&
\int \frac{d^3x\,d^3p}{(2\pi)^3} \left(E_s(x, \vec p) G_s(x, \vec p)
- \frac{E_q(x, \vec p)n_q(x, \vec p)}{\tau_{qs}(x, \vec p)} \right)
\\ \nonumber &+&
\int \frac{d^3x\,d^3p}{(2\pi)^3} \left(E_q(x, \vec p) G_q(x, \vec p)
- \frac{E_s(x, \vec p)n_s(x, \vec p)}{\tau_{sq}(x, \vec p)} \right)
\end{eqnarray}
for the change of the total energy due to the collisional part of
Eqs.~(\ref{relaxa}).  The factor $1/2$ at the left hand side accounts
for the antiparticle degree of freedom. Note that for the Lagrangian
(\ref{su3lag}) an effective potential, as it was given in
Eq.~(\ref{vexp}) for the $SU(2)$ case, exists in a mathematical sense,
however, due to the six fermion couplings in the 't~Hooft determinant,
it is not possible to compute it analytically. For the exact Boltzmann
equation (\ref{boltz}), each of the integrals on the right hand side of
Eq.~(\ref{bekons}) vanishes individually, even if the spatial
integrations are dropped. We adopt this stronger constraint also for
Eq.~(\ref{bekons}), and have the two conditions
\begin{mathletters} \label{tfit} \begin{eqnarray}
\int \frac{d^3p}{(2\pi)^3} \frac{E_q(x, \vec p)}{\tau_{qq}(x, \vec p)}
\left(\tilde n_q(x, \vec p) - n_q(x, \vec p) \right) &=& 0 \\
\int \frac{d^3p}{(2\pi)^3} \frac{E_s(x, \vec p)}{\tau_{ss}(x, \vec p)}
\left(\tilde n_s(x, \vec p) - n_s(x, \vec p) \right) &=& 0
\end{eqnarray} \end{mathletters}
to determine $T(x)$ and $\zeta(x)$ at each space--time point. In order
to fulfill the remaining constraints
\begin{mathletters} \label{gconst} \begin{eqnarray}
\int \frac{d^3p}{(2\pi)^3} \left(E_s(x, \vec p) G_s(x, \vec p)
- \frac{E_q(x, \vec p)n_q(x, \vec p)}{\tau_{qs}(x, \vec p)} \right) &=& 0 \\
\int \frac{d^3p}{(2\pi)^3} \left(E_q(x, \vec p) G_q(x, \vec p)
- \frac{E_s(x, \vec p)n_s(x, \vec p)}{\tau_{sq}(x, \vec p)} \right) &=& 0
\hspace{2mm},
\end{eqnarray} \end{mathletters}
we make the ansatz that the gain terms $G_q$ and $G_s$ are proportional
to the densities in local thermal equilibrium:
\begin{mathletters} \begin{eqnarray}
G_q(x, \vec p) &=&
\gamma_q(x) \frac{\tilde n_q(x, \vec p)}{\tau_{qs}(x, \vec p)} \\
G_s(x, \vec p) &=& \gamma_s(x) \frac{\tilde n_s(x, \vec p)/\zeta(x)}
{\tau_{sq}(x, \vec p)} \hspace{2mm}.
\end{eqnarray} \end{mathletters}
The factors $\gamma_q(x)$ and $\gamma_s(x)$ are determined by solving
Eqs.~(\ref{gconst}):
\begin{mathletters} \begin{eqnarray}
\gamma_q(x) &=& \int \frac{d^3p}{(2\pi)^3}
\frac{E_s(x, \vec p) n_s(x, \vec p)}{\tau_{sq}(x, \vec p)}
\Biggm/ \int \frac{d^3p}{(2\pi)^3}
\frac{E_q(x, \vec p) \tilde n_q(x, \vec p)}{\tau_{qs}(x, \vec p)} \\
\gamma_s(x) &=& \int \frac{d^3p}{(2\pi)^3}
\frac{E_q(x, \vec p) n_q(x, \vec p)}{\tau_{qs}(x, \vec p)}
\Biggm/ \int \frac{d^3p}{(2\pi)^3}
\frac{E_s(x, \vec p) \tilde n_s(x, \vec p)/\zeta(x)}{\tau_{sq}(x, \vec p)}
\hspace{2mm}.
\end{eqnarray} \end{mathletters}
With this form of the gain terms, the collision term vanishes in local
thermal equilibrium.

\subsection{Relaxation Times} \label{relsec}
The basic assumption of the relaxation time approximation is that the
system is close to equilibrium and that the right hand side of the
Boltzmann equation is only linear in the difference $n_f-\tilde n_f$.
Therefore the relaxation times have to be evaluated with the
equilibrium densities and masses and depend only on the local
temperature.

The collision term on the left hand side of Eq.~(\ref{boltz})
involves a sum over all possible scattering processes. The task of
computing the relaxation times for Eq.~(\ref{relaxa}) is thus twofold:
firstly one has to define the relaxation time for one single process
and afterwards add up the contributions of several processes in order
to obtain Eqs.~(\ref{relaxa}).

In order to define the the relaxation times, we go back to
Eq.~(\ref{boltz}) and write it in the form
\begin{equation}
\left(\partial_t + \vec v_f \vec \partial_x - \vec \partial_x E_f
\vec \partial_p \right) n_f(x, \vec p) =
\sum_P \left(I_{P, \rm gain}^f  - I_{P, \rm loss}^f\right)
\hspace{2mm},
\end{equation}
where the sum runs over all possible processes and $I_{P, \rm gain}^f$
and $I_{P, \rm loss}^f$ are the respective gain and loss terms due to
process $P$. We focus on one generic process $P:\, ff_1\leftrightarrow
f'f_1'$.  The loss term due to this process is given by
\begin{equation}
I_{P, \rm loss}^f = \int \frac{d^3p_1}{(2\pi)^3} d\Omega \,
v_{\rm rel} \frac{d\sigma_P}{d\Omega} n_f n_{f_1} \phi_{f'} \phi_{f_1'}
\hspace{2mm}.
\end{equation}
By comparing this to the ansatz (\ref{relaxa}), one immediately
obtains the relaxation time for this process
\begin{equation}
\frac{1}{\tau_P(x, \vec p)} = \int \frac{d^3p_1}{(2\pi)^3} d\Omega \,
v_{\rm rel} \frac{d\sigma_P}{d\Omega} n_{f_1} \phi_{f'} \phi_{f_1'}
\hspace{2mm}.
\end{equation}
In order to evaluate this expression, we go to the rest frame of
the plasma and approximate the differential cross section by its
equilibrium value at the local temperature $T(x)$, the
density of flavor $f_1$ by
\begin{equation}
n_{f_1}(T(x), p_1) =
\frac{2N_c}{\exp\left[E_{f_1}(T(x), p_1) / T(x)\right] + 1}
\hspace{2mm}, \label{fermi}
\end{equation}
where $E_{f_1}(T(x), p_1)$ is computed using the equilibrium mass
corresponding to the effective temperature $T(x)$. The blocking
factors are replaced by
\begin{equation}
\phi_{f'}(T(x), s) = 1 -
\frac{1}{\exp\left[E_{f'}^{\rm cm}(T(x), s) / T(x)\right] + 1} \hspace{2mm},
\end{equation}
where $E_{f'}^{\rm cm}(T(x), s)$ is the energy for the participant $f'$
in the centre of mass frame, which can be expressed as a function of
the masses and the Mandelstam variable $s$. These replacements lead to
\begin{equation}
\frac{1}{\tau_P(T(x), p_f)} = \int \frac{d^3p_1}{(2\pi)^3}
v_{\rm rel} \sigma_{P, \rm eff}(T(x), s) n_{f_1}(T(x), p_1)
\label{frankie} \hspace{2mm}.
\end{equation}
The effective cross section in Eq.~(\ref{frankie}) is defined by
\begin{equation}
\sigma_{P, \rm eff}(T(x), s) = \phi_{f'}(T(x), s) \phi_{f_1'}(T(x), s)
\int d\Omega \frac{d\sigma_P}{d\Omega} (T(x), s, t) \hspace{2mm}.
\end{equation}
By introducing the Mandelstam variable $s$ as integration variable,
Eq.~(\ref{frankie}) can be reduced to
\begin{eqnarray}
\frac{1}{\tau_P(T(x), p_f)} &=& \frac{1}{16\pi^2}\frac{1}{E_fp_f}
\int_{(m_f+m_{f_1})^2}^\infty ds \sqrt{[s-(m_f+m_{f_1})^2][s-(m_f-m_{f_1})^2]}
\nonumber \\ \label{relax} & & \hspace{3cm} \times
\sigma_{P, \rm eff}(T(x), s) \, \omega_{ff_1}(T(x), s, E_f)
\hspace{2mm},
\end{eqnarray}
where
\begin{equation}
\omega_{ff_1}(T(x), s, E_f)=\int dE_{f_1} n_{f_1}(T(x), p_1)
\Theta\left[\left(2p_fp_{f_1}\right)^2-
\left(s-m_f^2-m_{f_1}^2-2E_fE_{f_1}\right)^2\right]
\end{equation}
is a weight function, which can be computed analytically for $n_{f_1}$
given by Eq.~(\ref{fermi}). The temperature dependent cross sections
entering Eq.~(\ref{relax}) are taken from Ref.~\cite{su3elast}.  Note
that $\tau_P$ in the rest system of the plasma is a function of $T(x)$
and $p_f$ only. It can thus be easily obtained via table lookup. In the
lab system, we compute $\tau_P(x, \vec p)$ by applying a Lorentz boost
to Eq.~(\ref{relax}).

In order to relate the relaxation time for a specific process to those
needed in Eq.~(\ref{relaxa}), we consider again the loss term of the
exact equation (\ref{boltz}). To be specific, we set $f=u$. Then the
exact loss term can be decomposed into the contributions
\begin{eqnarray}
\sum_P I^u_{P,\rm loss} &=& I^u_{uu\to uu,\rm loss} + I^u_{ud\to ud,\rm loss}
+ I^u_{us\to us,\rm loss} \\ \nonumber
&+& I^u_{u\bar u\to u\bar u,\rm loss}
+ I^u_{u\bar d\to u\bar d,\rm loss} + I^u_{u\bar s\to u\bar s,\rm loss}
\\ \nonumber
&+& I^u_{u\bar u\to d\bar d,\rm loss} + I^u_{u\bar u\to s\bar s,\rm loss}
\hspace{2mm}.
\end{eqnarray}
Only the last two of these contributions correspond to inelastic
processes.  However, we have considered $u$ and $d$ to be degenerate in
Eq.~(\ref{relaxa}), so that we have to include also the contribution
of $u\bar u\to d\bar d$ to the elastic relaxation time $\tau_{qq}$ of
Eq.~(\ref{relaxa}). Thus one has
\begin{mathletters} \begin{eqnarray}
\frac{1}{\tau_{qq}} &=& \frac{1}{\tau_{uu\to uu}} + \frac{1}{\tau_{ud\to ud}}
+ \frac{\zeta}{\tau_{us\to us}} + \frac{1}{\tau_{u\bar u\to u\bar u}}
+ \frac{1}{\tau_{u\bar d\to u\bar d}} + \frac{\zeta}{\tau_{u\bar s\to u\bar s}}
+ \frac{1}{\tau_{u\bar u\to d\bar d}} \\
\frac{1}{\tau_{qs}} &=& \frac{1}{\tau_{u\bar u\to s\bar s}} \hspace{2mm},
\end{eqnarray} \end{mathletters}
where a factor $\zeta$ has been multiplied to the contributions of the
processes $us\to us$ and $u\bar s\to u\bar s$ in order to relate
Eq.~(\ref{fermi}) to Eq.~(\ref{zeta_und_mordio}). By considering the
loss term for the strange quark, one obtains in an analogous
fashion
\begin{mathletters} \begin{eqnarray}
\frac{1}{\tau_{ss}} &=& \frac{\zeta}{\tau_{ss\to ss}}
+ \frac{2}{\tau_{su\to su}} + \frac{\zeta}{\tau_{s\bar s\to s\bar s}}
+ \frac{2}{\tau_{s\bar u\to s\bar u}} \\
\frac{1}{\tau_{sq}} &=& \frac{2\zeta}{\tau_{s\bar s\to u\bar u}} \hspace{2mm},
\end{eqnarray} \end{mathletters}
where the factor 2 in the contributions of $su\to su$, $s\bar
u\to s\bar u$ and $s\bar s\to u\bar u$ accounts for the fact that in
all these processes $u$ can be replaced by $d$.

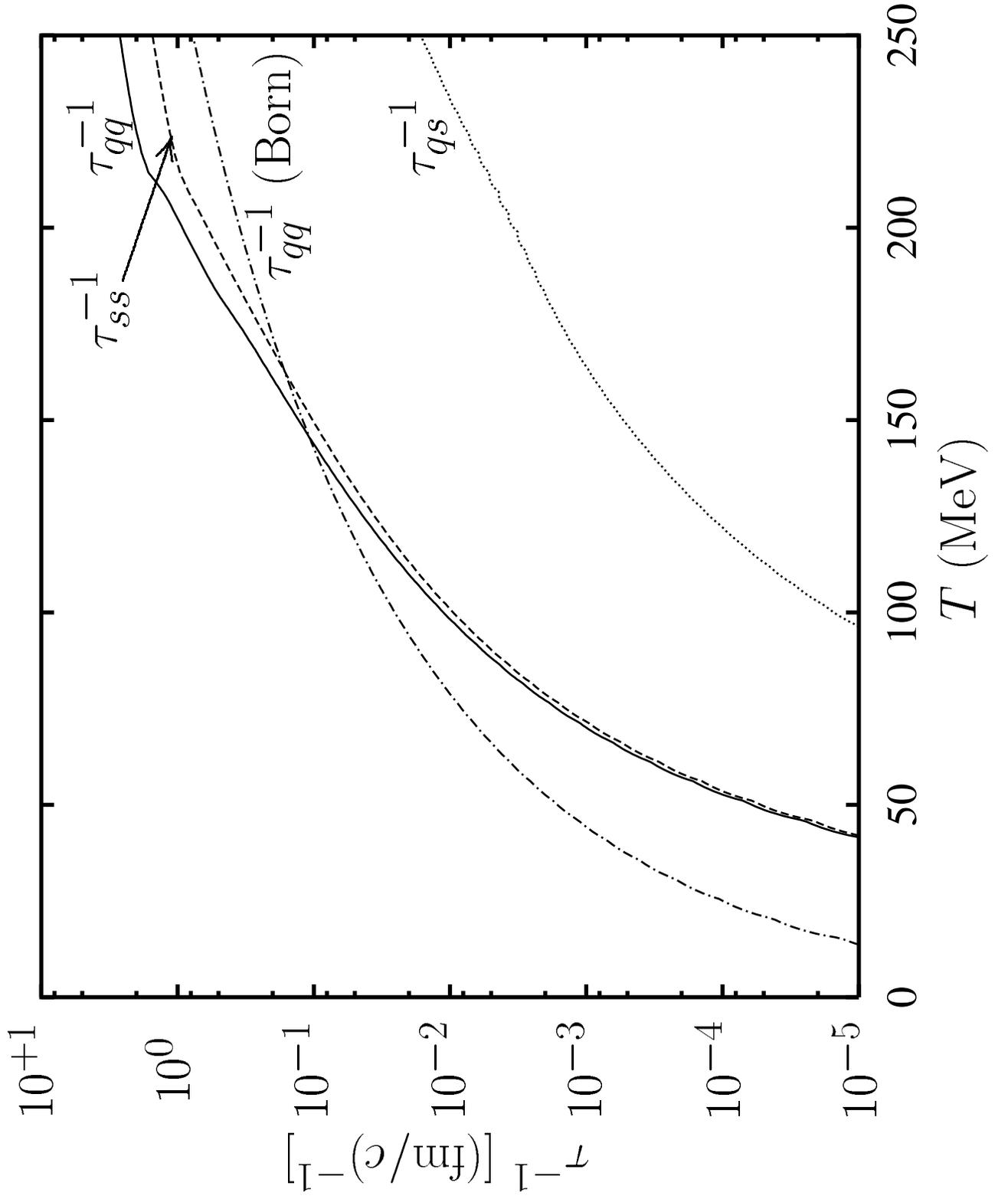
\begin{figure}
\caption[]{Inverse relaxation times $\tau_{qq}^{-1}$ (solid line),
	   $\tau_{ss}^{-1}$ (dashed line) and $\tau_{qs}^{-1}$ (dotted
	   line) as a function of temperature for $p=200$~MeV. The
           dot--dashed line shows $\tau_{qq}^{-1}$ in Born approximation.}
\label{relax200}
\end{figure}

In Fig.~\ref{relax200}, we show the resulting inverse relaxation times
$\tau^{-1}$ for the elastic scattering of light quarks (solid line),
the elastic scattering of strange quarks (dashed line) and for the
production of strange quarks (dotted line) as a function of temperature
and for a momentum $p=200$~MeV. While $\tau_{qq}^{-1}$ and
$\tau_{ss}^{-1}$ are approximately equal, $\tau_{qs}^{-1}$ is around
two orders of magnitude lower than $\tau_{qq}^{-1}$ for this value of
$p$.  The main reason for this is the threshold for the process $q\bar
q\to s\bar s$, which makes this process improbable at low momenta.  The
ratio between $\tau_{qq}^{-1}$ and $\tau_{qs}^{-1}$ decreases for
increasing momenta. At low temperatures, $\tau^{-1}$ goes to zero
rapidly, mainly because of the density factor in Eq.~(\ref{frankie}).
The solid, dashed and dotted curves of Fig.~\ref{relax200} are
calculated using temperature dependent quark masses and the temperature
dependent cross sections given in Ref.~\cite{su3elast}. These are to be
contrasted with the dot--dashed line, which gives $\tau_{qq}^{-1}$ from
a calculation, where the masses have been set to the current quark
masses and the cross sections have been calculated in the Born
approximation \cite{su2elast}, which does not include any medium
dependence. At high temperatures, this procedure results in an inverse
relaxation time, which is approximately a factor 3.5 smaller, whereas
the falloff at low temperatures is weaker due to the lower masses.

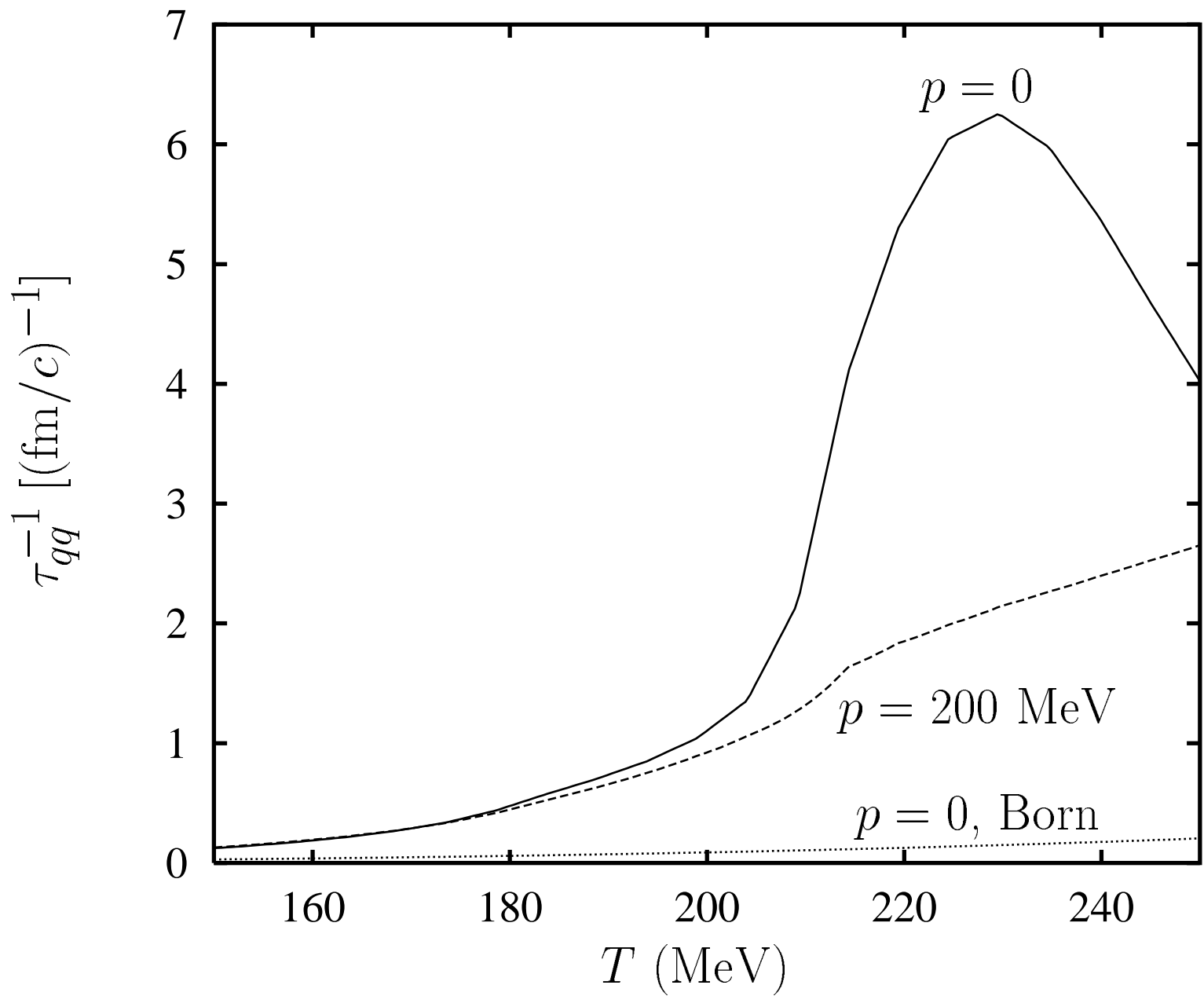
\begin{figure}
\caption[]{Inverse relaxation time $\tau_{qq}^{-1}$ as a function of
           temperature for $p=0$ (solid line), $p=200$~MeV (dashed line)
           and at $p=0$ in Born approximation (dotted line).}
\label{relax000}
\end{figure}

In Fig.~\ref{relax000}, $\tau_{qq}^{-1}$ is given as a function of
temperature for $p=0$ (solid line) and $p=200$~MeV (dashed line). One
notes that the solid line displays a pronounced maximum at $T\approx
230$~MeV.  It originates from a divergence of the scattering length,
which occurs for the quark--antiquark scattering processes at the Mott
temperature, where the pion becomes unbound \cite{su3elast,gerry}. This
phenomenon is related to critical scattering. The Mott temperature
is ca.~210~MeV for our parameter set. As to be expected, the Born
calculation for $p=0$, which is given by the dotted line of
Fig.~\ref{relax000}, displays no structure at this temperature.

\subsection{Numerical Results}
\subsubsection{Densities, Quark Masses, Spectra and Strangeness Yield}
The transport equations (\ref{relaxa}) are solved using an obvious
generalization of the numerical methods outlined in
Section~\ref{nummet}. In this subsection we show results of a
calculation with the initial condition
\begin{mathletters} \label{su3ini} \begin{eqnarray}
n_q(x, \vec p) &=& \exp\left(-\frac{r^2}{2r_0^2}\right)
                 \frac{4N_c}{\exp\left(\sqrt{\vec p^2+m_i^2}/T_0\right)+1}
                 \label{su3inia} \\
n_s(x, \vec p) &=& 0 \hspace{2mm}.
\end{eqnarray} \end{mathletters}
Comparing the form of Eq.~(\ref{su3inia}) with Eq.~(\ref{inicon}) there
is only one difference, namely the absence of a cutoff in momentum
space: While Eq.~(\ref{inicon}) contains a {\em smooth\/} cutoff, we
now continue the Fermi shape of the distribution up to the grid
boundary $p_{\rm max}$, which in our standard parameter set is set to
$p_{\rm max}=\Lambda$. This modification is necessary if one wants to
determine effective temperatures from an exponential fit to the
particle spectra.  The disadvantage of this modification is that the
total particle number and the total energy are no longer exactly
conserved. We will discuss the magnitude of these effects later.

The standard values of the parameters are the same as used in
Section~\ref{vnumres}: $r_0=3$~fm and $T_0=240$~MeV; the initialization
mass $m_i$ is taken to be a solution of Eq.~(\ref{su3gap}) at $r=0$. The
NJL model parameters for all $SU_f(3)$ calculations are taken from
Ref.~\cite{su3hadron} and are $m_{0q}=5.5$~MeV, $m_{0s}=140.7$~MeV,
$G\Lambda^2=1.835$, $K\Lambda^5=12.36$ and $\Lambda=602.3$~MeV.

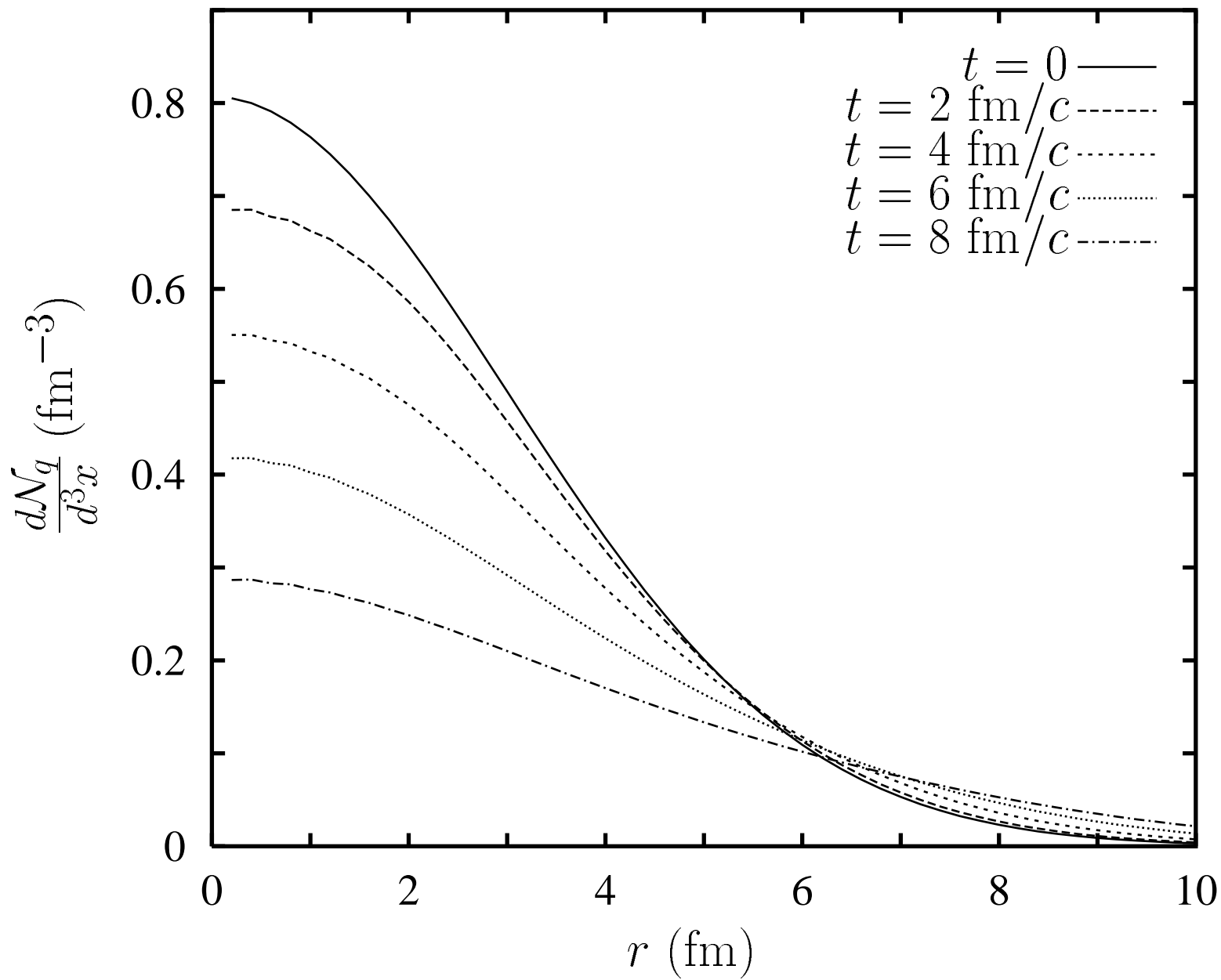
\begin{figure}
\caption[]{Light quark density at the same times as in Fig.~\ref{qdens}
	   for the full calculation containing three flavors and
	   collisions.}
\label{bqdens}
\end{figure}

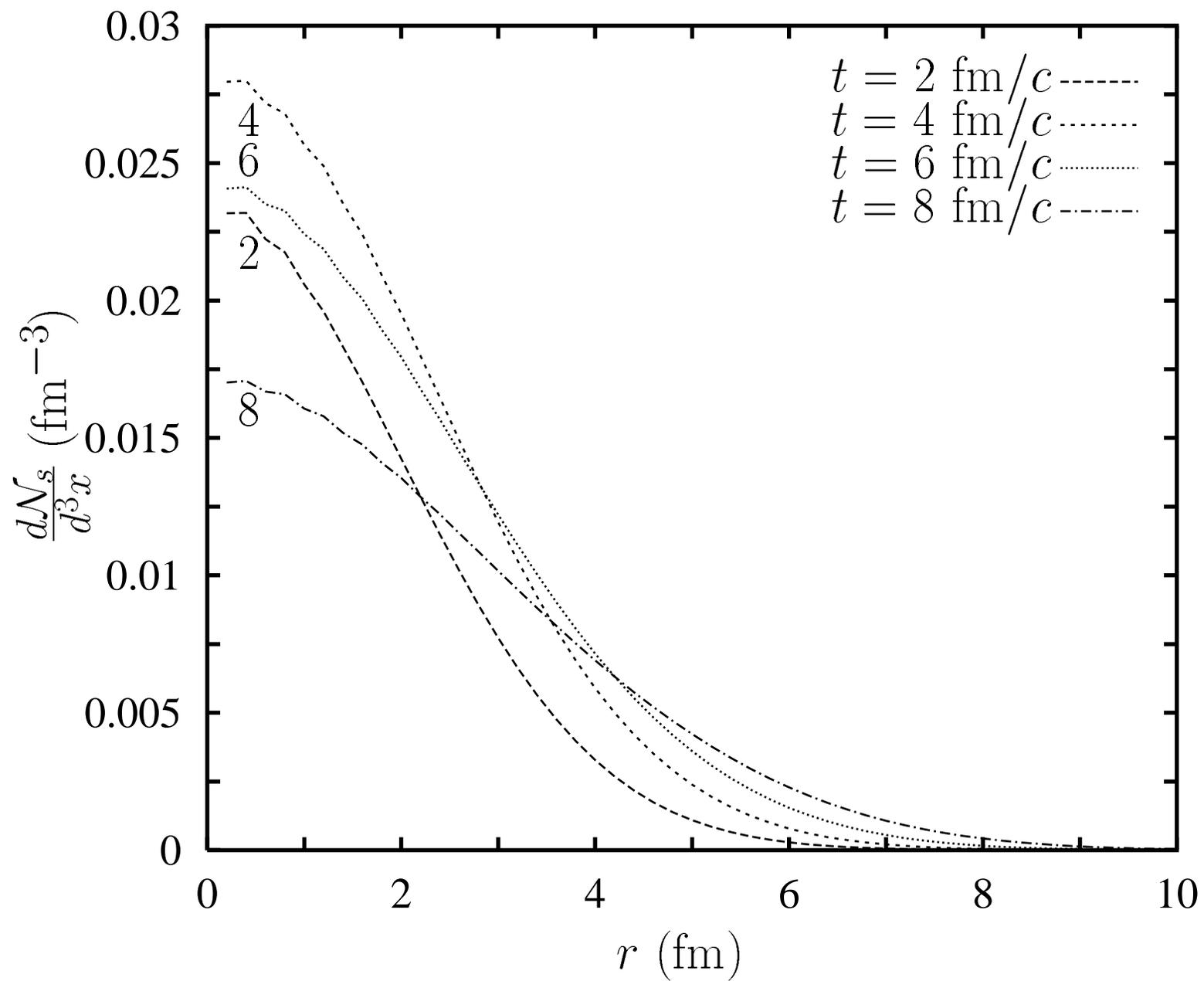
\begin{figure}
\caption[]{Strange quark density for the same times as in
	   Fig.~\ref{qdens} for the full calculation containing three
	   flavors and collisions. Note that the density is zero at
	   $t=0$ and rises until $t=4$~fm$/c$ due to production by
	   the process $q\bar q\to s\bar s$.}
\label{sdens}
\end{figure}

The time dependence of the light quark density is shown in
Fig.~\ref{bqdens}. This figure should be compared with the same
quantity calculated for the Vlasov equation, shown in
Fig.~\ref{qdens}.  The time steps coincide in both figures. The
behaviour of the light quark densities is rather similar. The time
variation for the density of strange quarks is shown in
Fig.~\ref{sdens}. When comparing it to Fig.~\ref{bqdens} for the light
quarks, one should first note the different scale on the ordinate
(about a factor of 30 down), and secondly the initial condition $n_s(t =
0, \vec x, \vec p) = 0$. The density {\em rises\/} due to strange quark
production until $t=4$~fm$/c$, after which time the rarefaction due to
flow overcomes the production and the density drops.

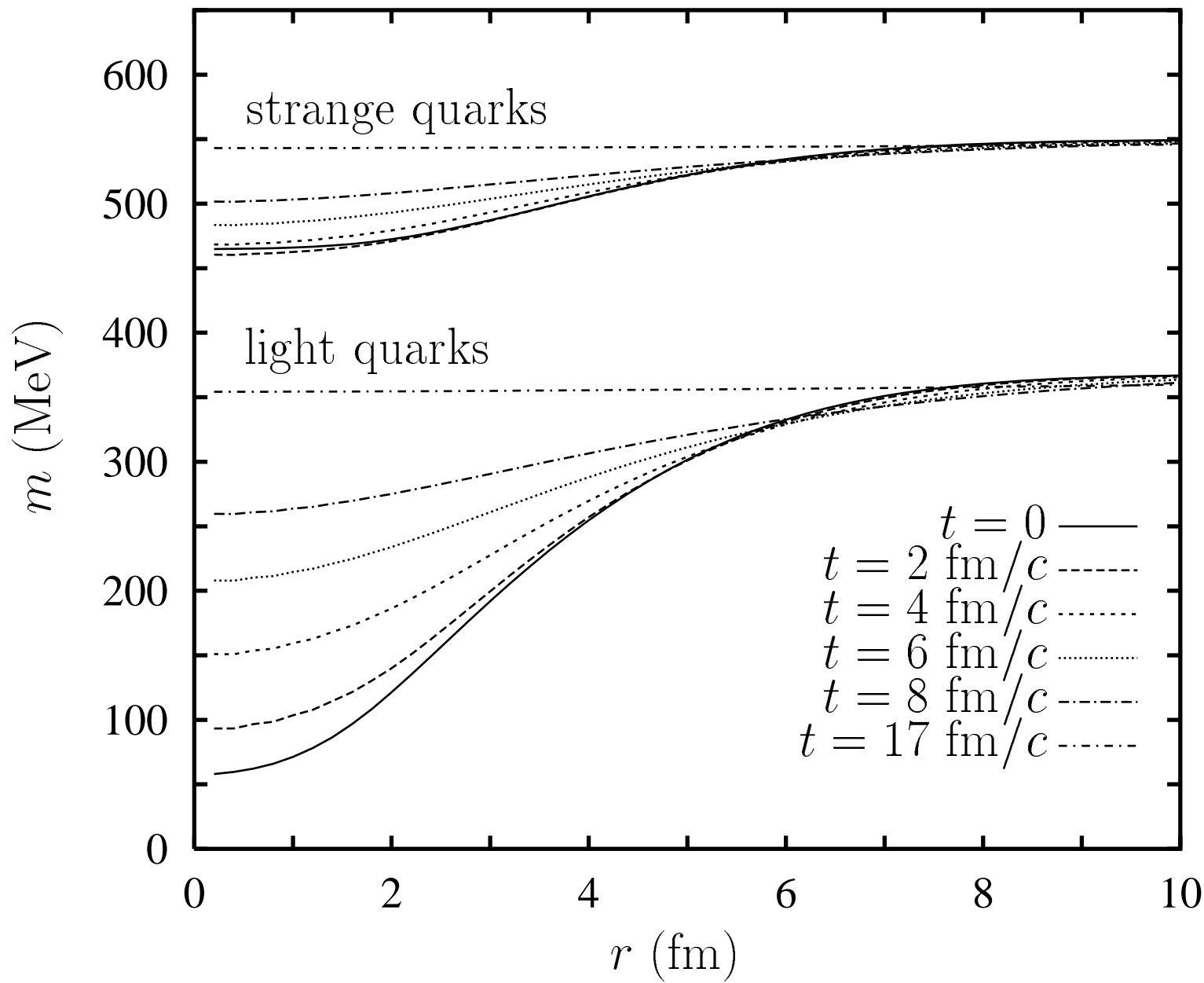
\begin{figure}
\caption[]{Light and strange quark masses as a function of time.}
\label{bqmass}
\end{figure}

The light and strange quark masses are shown in Fig.~\ref{bqmass} as a
function of time. For the light quark mass one obtains essentially the
same behaviour as observed in the $SU(2)$ case shown in
Fig.~\ref{qmass}. At $t=0$ the light quark mass amounts to $m_q=58$~MeV
in the centre and rises to the vacuum value $m_q=368$~MeV in the outer
regions. As time progresses, the masses increase everywhere to this
value. For the strange quarks, on the other hand, one does not observe
strong variations of the constituent mass.  This different behaviour in
the space-time behavior of $m_q(\vec x, t)$ and $m_s(\vec x, t)$ is
significant.  While the difference in the vacuum values
$m_s-m_q=182$~MeV compares well with the mass difference between lambda
and proton, $m_\Lambda - m_p=177$~MeV, the in medium corrections in the
plasma are large for the light quarks and small for the strange quarks.
In the NJL model the medium effects on the masses can be traced to the
Pauli-blocking, essentially. The Pauli-blocking for the strange quarks
is small because of the low density of this kind of quarks.  At $t=0$,
the strange quark density vanishes and only the term proportional to
$K$ in Eq.~(\ref{su3gap}) gives corrections to the vacuum mass. This
leads to a value of $m_s=465$~MeV in the centre and $m_s=550$~MeV in
the outer regions. At later times, the term proportional to $G$ also
gives medium corrections. However, these are not sufficient to lower
the strange quark mass substantially in the central region. With
increasing time, also the strange quark mass evolves to its vacuum
expectation value everywhere, as expected.  The large mass difference
of the in medium masses suppresses the production of strange quarks
in our calculation -- and possibly in nature.

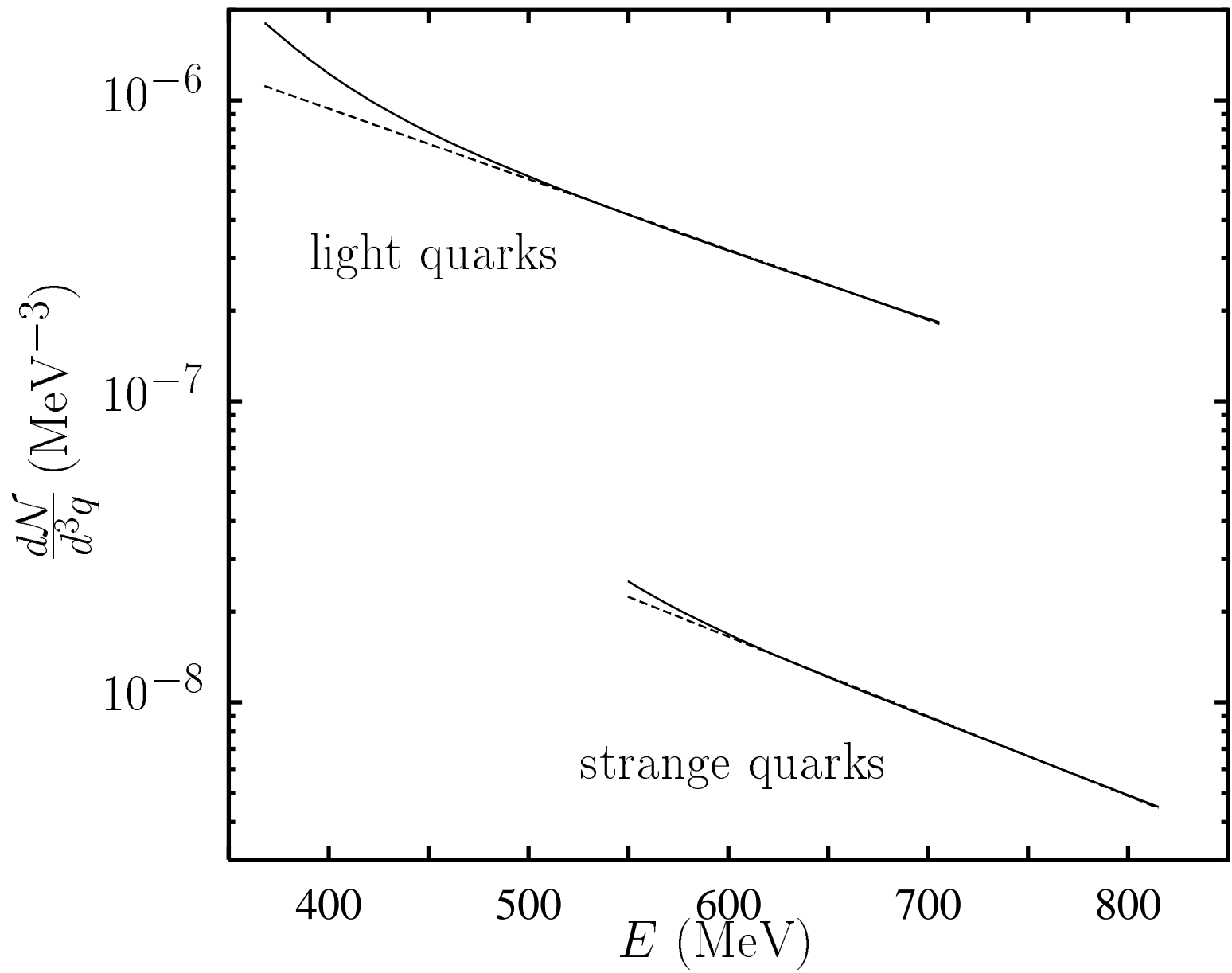
\begin{figure}
\caption[]{Distributions of light and strange quarks
	   for the final state as a function of the particle energy
	   (solid lines). The dashed lines are exponential fits with
	   the slope temperatures $T_q=185$~MeV for light quarks and
	   $T_s=164$~MeV for strange quarks.}
\label{bspekt}
\end{figure}

In Fig.~\ref{bspekt}, we show the ``observed'' (for $t\to\infty$)
distributions of particles in momentum space, $d{\cal N} / d^3p$, as a
function of the particle energy. An exponential fit is applied to the
high energy part of the spectra (dashed lines). The effective slope
temperatures of these fits for the light and strange quarks are
different: $T_q=185$~MeV for light quarks and $T_s=164$~MeV for the
strange ones. The calculated spectra show an enhancement at low
energies compared to the exponential fit.  This enhancement arises from
the mean field, as has already been explained in Section~\ref{vnumres}.
We comment on the difference $T_q-T_s$ in Section~\ref{etranger}.

We find for the strangeness yield, defined as the ratio of the number
of strange quarks to that of light quarks at $t=\infty$, a value of
${\cal N}_{s,t=\infty} / {\cal N}_{q,t=\infty} = 2.2\%$. This value is
rather low and is mainly determined by incomplete chemical
equilibration, as we will show in in Section~\ref{etranger}.

\subsubsection{Dependencies on the Grid Settings and Numerical Accuracy}
In order to see the reliability of the results and the stability with
respect to the numerical scheme, we have investigated the dependence of
the particle multiplicities and the spectra on the grid extension.  The
results of the various runs are compiled in Table~\ref{multitab}. The
calculated multiplicities of the light and strange quarks with the
standard parameters described above are given in the row labeled (a).
One has ${\cal N}_{q,t=0}=343.2$ in the initial state and ${\cal
N}_{q,t=\infty}=365.6$ and ${\cal N}_{s,t=\infty}=8.1$ in the final
state.  This corresponds to a relative yield of ${\cal N}_{s,t=\infty}
/ {\cal N}_{q,t=\infty} = 2.2\%$. We observe an increase of the total
multiplicity from $\left({\cal N}_q+{\cal N}_s\right)_{t=0}=343.2$ to
$\left({\cal N}_q+{\cal N}_s\right)_{t=\infty}=373.7$. However, ${\cal
N}_q+{\cal N}_s$ should be a constant of motion for the exact Boltzmann
equation (\ref{boltz}).

The possible reasons for the increase of the increase are the
following: (i) numerical inaccuracies, (ii) the relaxation time ansatz,
which, however, should lead to a decrease of the particle number due to
the dropping temperature and (iii) the extension of the grid in
momentum space. The last effect can be understood as follows:
In the calculation related to row (a) of Table~\ref{multitab}, the grid
has maximal momentum $p_{\rm max}=\Lambda=602$~MeV and the initial
``thermal'' distribution of Eqs.~(\ref{su3ini}) is cut off at this
value.  Calculating also the particles from Eq.~(\ref{su3ini}) with
$p\ge p_{\rm max}$, we see that our grid includes only 40\% of all
particles {\em explicitly\/}.  Nevertheless, due to the open boundary
conditions which are chosen in Eq.~(\ref{discrete}), i.\,e. smoothness
at $p=p_{\rm max}$, the numerical procedure assumes that the particle
density can be extrapolated analytically {\em beyond\/} the grid
boundary and thus ``recognizes'' the remaining particles {\em
implicitly\/}, at least as far as the derivative terms of the transport
equation are concerned. Since the mean field tends to slow down
particles, it is possible that particles with $p\ge p_{\rm max}$, which
are not counted at $t=0$, appear inside the grid at later times, and
thus the multiplicity increases.

In order to eliminate the influence of the finite grid extension in
momentum space, two other settings have been used: firstly, the grid is
extended to include momenta up to $p_{\rm max}=3\Lambda$.  In this
calculation, all momentum integrals have been extended to $p=p_{\rm
max}$, except those occurring in the gap equation (\ref{su3gap}).  The
resulting multiplicities are given in row (b) of Table~\ref{multitab}.
Compared to the 8.9\% increase observed in the previous calculation,
one now obtains a increase of 5.8\% of the particle number, which is
comparable with the Vlasov result.  The fraction of strange particles
in this calculation is 3.9\%, compared to 2.2\% for calculation (a).

In a second investigation a smooth cutoff for the high momentum region
is introduced in the initial conditions, as is done for the Vlasov
equation in Eq.~(\ref{inicon}). In this calculation, the effective
distribution functions of Eqs.~(\ref{neff}) are consistently cut off by
the same factor. This procedure is also consistent with the requirement
of energy conservation, as has been explained in Section \ref{conserv},
and with the nature of the NJL model as a low energy theory.  The
multiplicities for this calculation are given in row (c) of
Table~\ref{multitab}. We observe a 1.3\% decrease of the particle
number, which means that the particle number is in good approximation
conserved. The strangeness content in the final state amounts to
2.3\%.

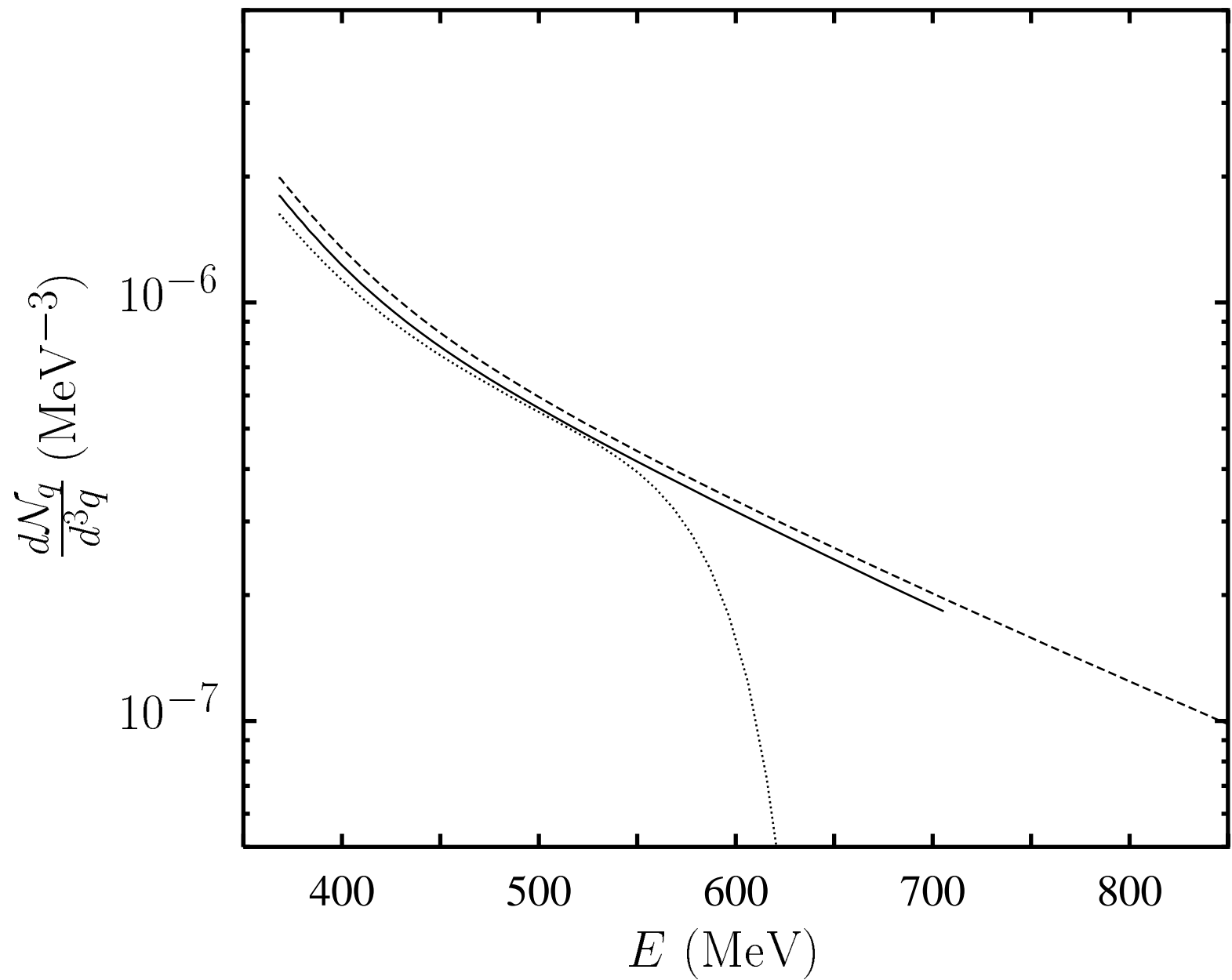
\begin{figure}
\caption[]{Distributions of light quarks resulting from
           different treatments of the high momentum part. Solid line:
           only particles with momentum $p\le\Lambda$ are included.
           Dashed line: particles with momentum $p\le3\Lambda$ are
           included. Dotted line: smooth cutoff of high momentum
           particles.}
\label{varspekt}
\end{figure}

The light quark spectra of calculations (a), (b) and (c) are shown in
Fig.~\ref{varspekt}. The data for the calculation (a) (solid line)
agree with those for calculation (b) (dashed line). The data for the
calculation (c) are given by the dotted line. Although calculation (c)
agrees reasonably with (a) and (b) in the low momentum region (as it
should), the momentum cutoff factor distorts the high momentum part
completely.  It is thus impossible to fit a reasonable slope
temperature to this curve.

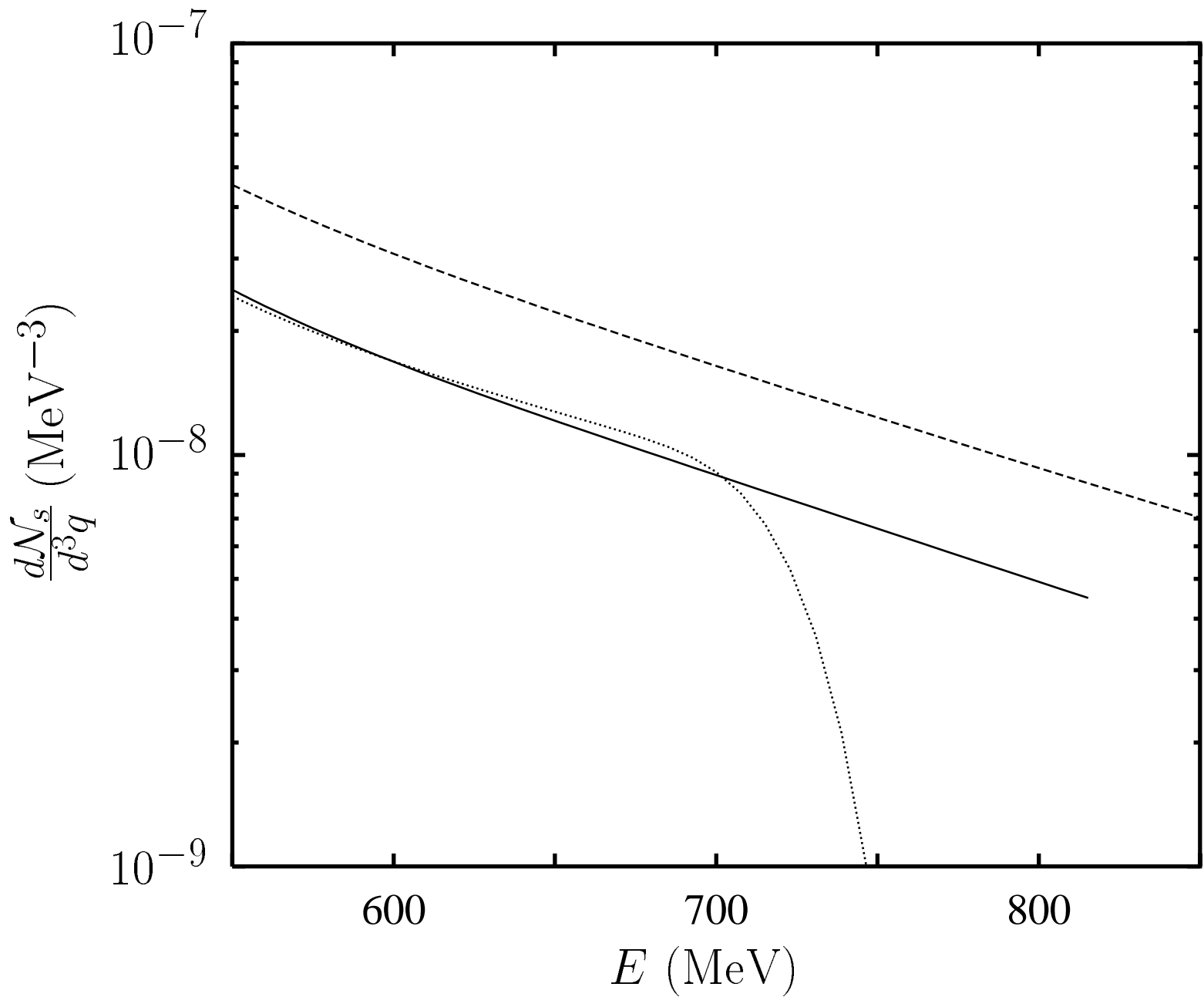
\begin{figure}
\caption[]{Distributions of strange quarks resulting from
           different treatments of the high momentum part. Solid line:
           only particles with momentum $p\le\Lambda$ are included.
           Dashed line: particles with momentum $p\le3\Lambda$ are
           included. Dotted line: smooth cutoff of high momentum
           particles.}
\label{vasspekt}
\end{figure}

While the agreement of the spectra is good for light quarks, this is
not the case for strange quarks, as can be seen from
Fig.~\ref{vasspekt}. Calculation (b), i.\,e. $p_{\rm max}=3\Lambda$,
gives a twice higher value for $d{\cal N}_s/d^3p$ as compared to
calculation (a) and (c), i.\,e. $p_{\rm max}=\Lambda$, since more light
quarks with high energy are available for the production of strange
quarks.

This may be the appropriate place to report the numerical effort and
accuracy.  The use of a partial differential equation in four variables
for the expansion of a spherical system in phase space is limited by
the use of computer memory and CPU time. For a $50^3$ grid, our program
uses ca.  3~MByte memory and 20~seconds of CPU time per time slice on a
Sparc~20 workstation. The grid size as well as the number of time
slices has to be increased when e.\,g. increasing the radius $r_0$ of
the fireball, so that the above numbers have to be scaled accordingly.
Most of the CPU time is consumed by the determination of the effective
temperature.  All indicators point to an overall {\em numerical\/}
accuracy of a few percent of our results.

\subsubsection{Dependencies on Medium Effects}
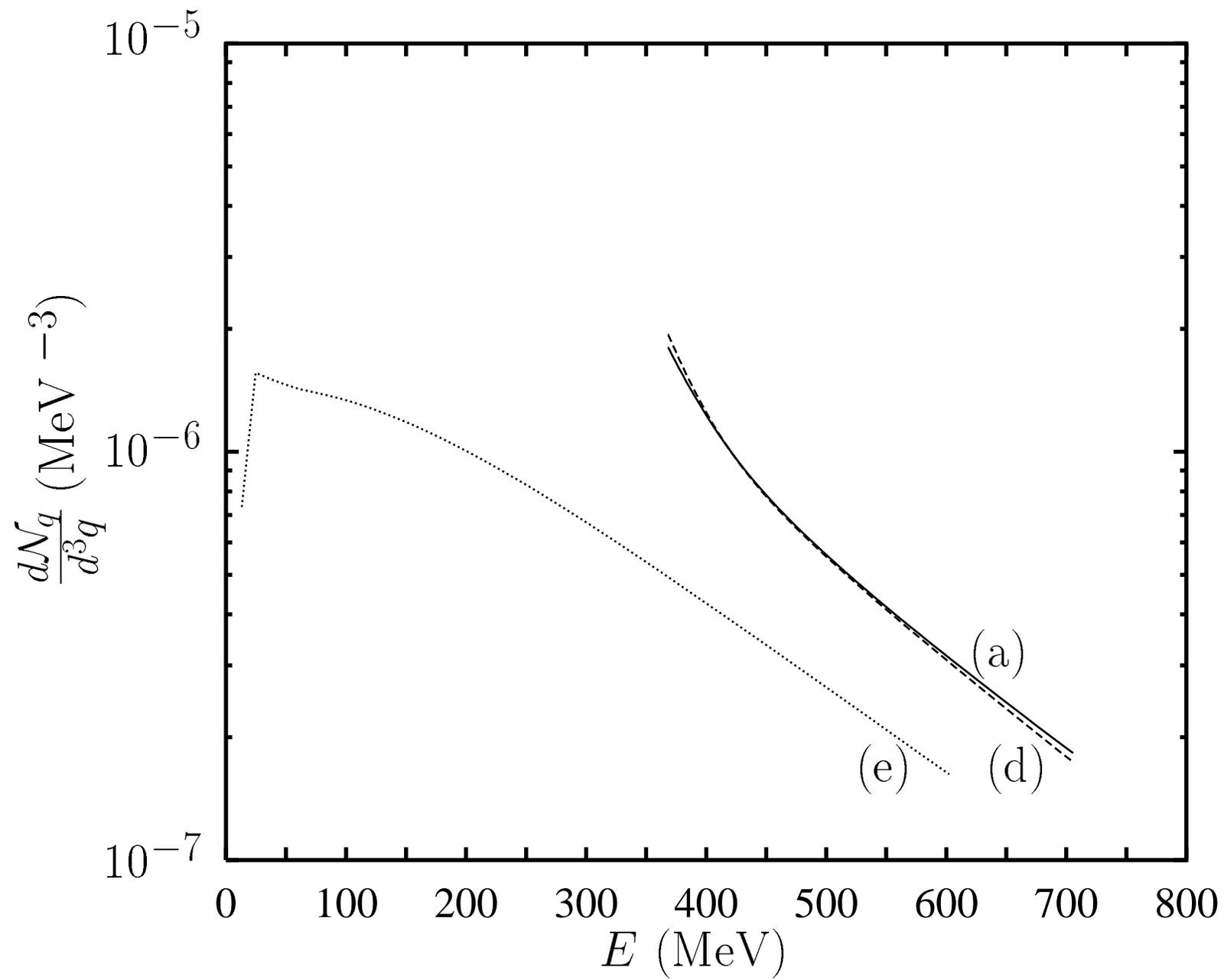
\begin{figure}
\caption[]{Light quark spectra for three different treatments of medium
	   effects; (a): standard procedure (solid line), (d): cross
	   sections in Born approximation (dashed line) and (e): cross
	   sections in Born approximation and mean field switched off.}
\label{bornspekt}
\end{figure}

The inclusion of medium effects in the collision terms is an important
and new aspect of our calculation. Indeed, the medium effects are
large, both for the relaxation times shown in Fig.~\ref{relax000}, as
well as for the quark masses shown in Fig.~\ref{bqmass}. It is thus
important to investigate their influence on the final spectra.  Our
results are compiled in Table~\ref{borntab}, where row (a) relates to
the standard calculation. The medium effects are studied in two ways:
by keeping or switching off the mean field and by replacing the cross
sections in the relaxation times by their Born approximation
\cite{su2elast} values.  All other parameters correspond to the
standard set.  Multiplicities and slope temperatures for these runs are
given in rows (d) and (e) of Table~\ref{borntab}, respectively.  While
the results of calculation (d) are similar to those of calculation (a),
calculation (e), i.\,e.  the one without mean field, gives completely
different results for multiplicities and slope temperatures. This can
also be seen in the light quark spectra given in Fig.~\ref{bornspekt}
where the results of calculations (a), (d) and (e) are shown by the
solid, dashed and dotted line, respectively. While (a) and (d) give
almost identical results, calculation (e) differs substantially.
Firstly, the energy range is shifted to smaller energies due to the
small quark mass. Secondly, the spectrum is flatter, indicating a
higher slope temperature. Thirdly, calculation (e) shows a deficiency
of particles at low energies, instead of an enhancement like (a) and
(d). This can be traced back to the fact that there is no mean field
any longer, which shifts particles into this region.

We have also investigated, whether the maximum in the temperature
dependence of the relaxation times in Fig.~\ref{relax000}, i.\,e.
critical scattering, has some effect on the final spectra. To this
purpose we have artificially set $\tau(p, T)=\tau(p=200~{\rm MeV}, T)$
for $p<200$~MeV and have compared the result with the full equation.
The resulting spectra agree with those of Fig.~\ref{bspekt} within the
accuracy of our calculation.

We conclude that (i) medium effects in the collision terms have small
effects on the shape of the final spectra but they considerably change
the strangeness yield (here by 35\%) and (ii) the medium effects in the
mean field parts lead to significant effects in the spectra, especially
in the low energy part.

\section{Strangeness Production} \label{etranger}
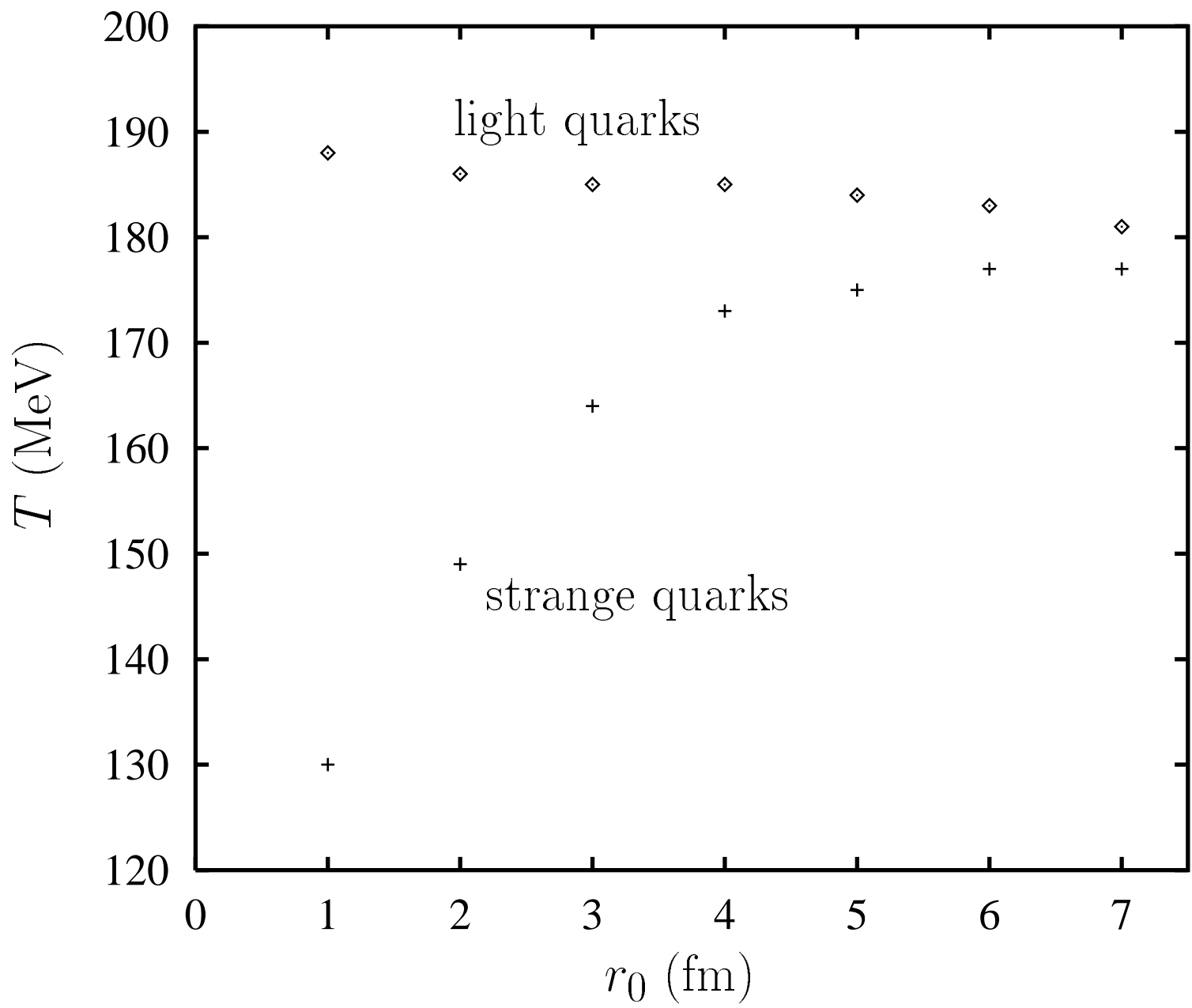
\begin{figure}
\caption[]{Slope temperatures for light quarks (diamonds) and strange
           quarks (crosses) as a function of the radius $r_0$ initial
           fireball.}
\label{slope_r}
\end{figure}

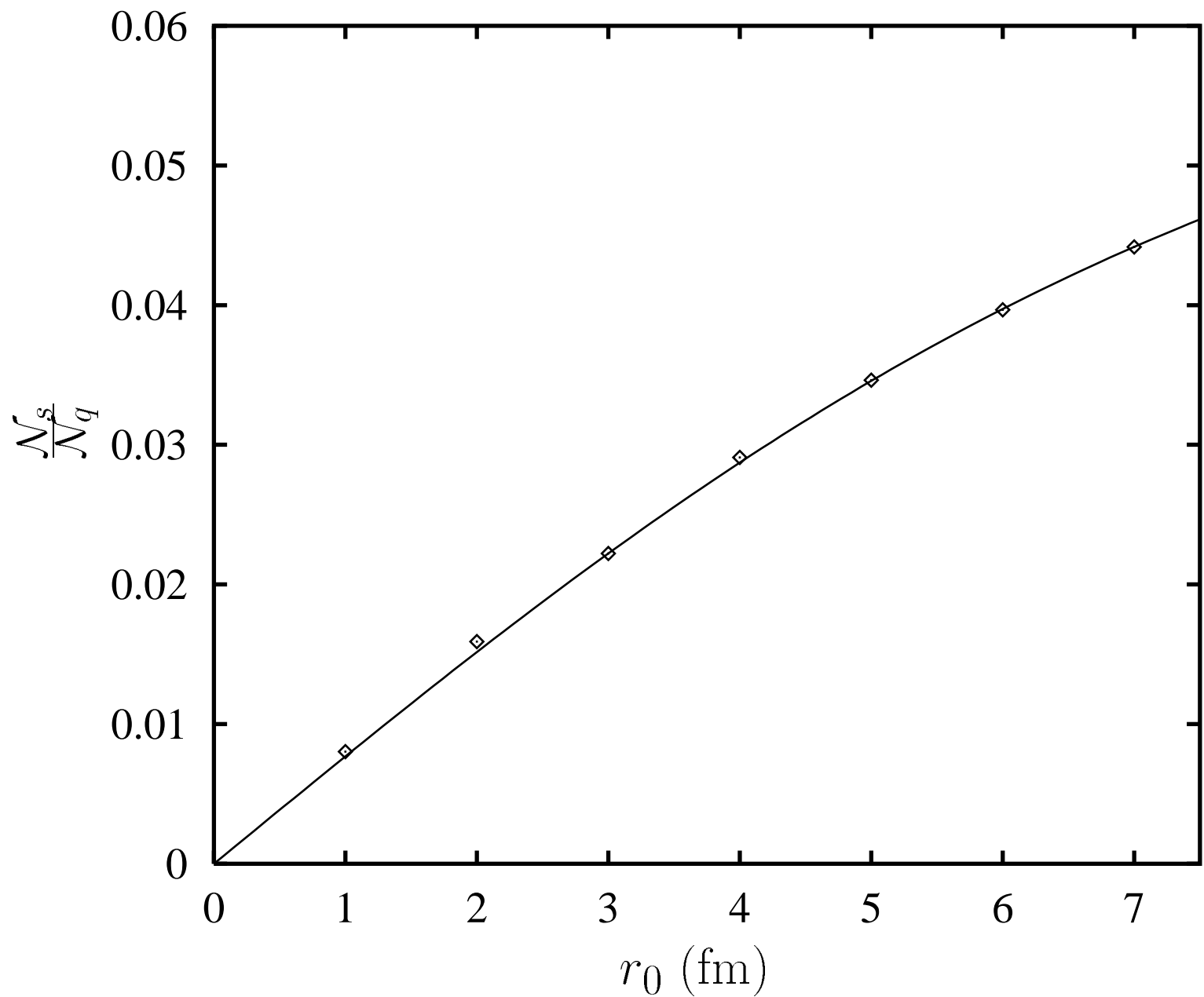
\begin{figure}
\caption[]{Variation of the strangeness yield as a function of the
	   initial radius (diamonds). The solid line is a fit by a
	   hyperbolic tangent.}
\label{tanhfit}
\end{figure}

We discuss the variation of the strangeness yield with the radius $r_0$
of the initial fireball.  The observables of the final state are the
slope temperatures $T_q$ and $T_s$ for the quarks and the strangeness
yields.  We have gradually increased the initial radius from $r_0=1$~fm
to $r_0=7$~fm while the other parameters of Eq.~(\ref{su3ini}) are
kept fixed. The resulting values for the slope temperatures $T_q$ and
$T_s$ and the strangeness yield ${\cal N}_s/{\cal N}_q$ are given in
Figs.~\ref{slope_r} and \ref{tanhfit}.  While the slope temperature for
light quarks hardly changes with the system size, the slope temperature
for the strange quarks rises and approaches $T_q$ from below.

The independence of $T_q$ on $r_0$ corresponds to the independence of
the hadronic slope temperatures on the projectile and target size,
which has been reported in Ref.~\cite{herr}. The fact that $T_s$ stays
below $T_q$ seems to be in contradiction with experimental data
\cite{NA44} and hydrodynamical models \cite{csolo}.  These models claim
that the slope temperatures scale with the particle mass like
$T=T_0+m\left<v_\bot\right>^2$, where $\left<v_\bot\right>$ is the mean
transverse expansion velocity. One has to keep in mind, however, that
this theoretical result is significantly influenced by the
cylindrical geometry of the colliding nuclei and by the presence of
transverse flow. These features are not present in our calculation.

The increase of $T_s$ with the initial radius $r_0$, which can be seen
in Fig.~\ref{slope_r}, is essentially an effect of thermal
equilibration. For small initial size, where there is negligible
thermal equilibration, one observes the ``temperature'' of the strange
quarks directly from the creation process. Its low value can be
understood {\em qualitatively\/} by assuming that the mean energy per
particle $E=m_q+3/2 T_q= m_s + 3/2 T_s$ is conserved during the
creation of a strange quark pair.  One thus obtains
$T_s=T_q-2/3(m_s-m_q)$ immediately after the pair creation. Subsequent
elastic collisions, as they will occur more frequently in the larger
systems, result in $T_s$ approaching $T_q$.  This is indeed observed in
Fig.~\ref{slope_r}.

According to Fig.~\ref{tanhfit}, the strangeness yield grows with the
radius of the initial fireball. This behaviour of the curve can be
understood using the following simple analytical model. We make the
ansatz
\begin{equation} \label{sansatz}
\frac{d\rho_s(\vec x, t)}{dt}
+ {\rm div}\left[\vec v \rho_s(\vec x, t)\right]
= \alpha \left[\rho_q^2(\vec x, t)
- \kappa \rho_s^2(\vec x, t) \right]
\end{equation}
for the space-time evolution of the strange quark density. In
Eq.~(\ref{sansatz}), the densities are parametrized by
\begin{mathletters} \label{par1} \begin{eqnarray}
\rho_q(\vec x, t) &=& \frac{{\cal N}_q}{\frac{4\pi}{3}r_q^3(t)}
\Theta(r_q(t) - |\vec x|) \\
\rho_s(\vec x, t) &=& \frac{{\cal N}_s(t)}{\frac{4\pi}{3}r_s^3(t)}
\Theta(r_s(t) - |\vec x|) \hspace{2mm}.
\end{eqnarray} \end{mathletters}
For the radius of the light quark system, we make the ansatz
\begin{equation} \label{rqt}
r_q^2(t) = r_0^2 + v^2t^2
\end{equation}
with a constant velocity $v$. Equation~(\ref{rqt}) is motivated by
the exact solution of the Vlasov equation in the noninteracting case.
We assume that the radius of the strange quark system
behaves similarly, however, is smaller by a factor $z$:
\begin{equation} \label{rst}
r_s(t) = z r_q(t) \hspace{2mm}.
\end{equation}
This factor accounts for the fact that only in the interior of the
fireball the density is large enough to produce particles and is also
born out numerically by comparing Figs.~\ref{bqdens} and \ref{sdens}.
With the parametrizations (\ref{par1})--(\ref{rst}) at hand,
Eq.~(\ref{sansatz}) can be solved analytically to give
\begin{equation}
\frac{{\cal N}_s(t)}{{\cal N}_q} = \sqrt{\frac{z^3}{\kappa}}
\tanh\left(\frac{3\alpha\sqrt{\kappa}{\cal N}_q}{4\pi r_0^2}
\frac{t}{\sqrt{r_0^2+v^2t^2}}\right) \hspace{2mm}.
\end{equation}
In the limit $t\to\infty$, one thus has
\begin{equation} \label{sresult}
\frac{{\cal N}_s}{{\cal N}_q} = \sqrt{\frac{z^3}{\kappa}}
\tanh\left(\frac{3\alpha\sqrt{\kappa}{\cal N}_q}{4\pi r_0^2v}
\right) \hspace{2mm}.
\end{equation}
Since ${\cal N}_q\sim r_0^3$, Eq.~(\ref{sresult}) predicts a dependence
on the initial radius as
\begin{equation}
\frac{{\cal N}_s}{{\cal N}_q} \sim \tanh(a r_0) \hspace{2mm}. \label{tanh1}
\end{equation}
Figure~\ref{tanhfit} shows, that this behaviour is indeed fulfilled in
our calculation. The fit shown in Fig.~\ref{tanhfit} (solid line) has
the parameters
\begin{equation}
\frac{{\cal N}_s}{{\cal N}_q} = 0.065 \tanh(0.12 r_0) \hspace{2mm}.
\label{tanh2}
\end{equation}
The factor 6.5\% in front of the hyperbolic tangent is the equilibrium
value for an infinite system.  A similar dependence like
Eq.~(\ref{tanh1}), however with different fit coefficients, can be seen
in the data for strangeness multiplicities at constant beam energy and
variing mass number \cite{marek}. The reason for the behaviour in
Eq.~(\ref{tanh1}) is that for larger systems the interior zone keeps
together longer and thus comes closer to chemical equilibrium.

The asymptotic value of 6.5\% in Eq.~(\ref{tanh2}) is about a factor of
three small than the data given in Ref.~\cite{marek}.
There may be several reasons for this discrepancy. The first one is
geometrical: Equation~\ref{sresult} contains a factor $z^{3/2}$, which
accounts for the smaller size of the strange system compared to the
light quark system. Estimating this factor to be $z=2/3$ from our
numerical results, this suppresses the strangeness yield by a factor
two. It is probable that this factor will appear with a different power
for a cylindrically symmetric system. The second reason for our low
strangeness yield is that strangeness production in the initial high
energy nucleon--nucleon collisions. like $pp\to p\Lambda K$, and
strangeness production in the hadronic phase like $\pi\rho\to KK$ are
absent from our calculation.

\section{Summary and Conclusions} \label{unconclusive}
We have developed numerical methods for the solution of transport
equations whose physics is derived from the two and three flavor NJL
Lagrangian and have given explicit solutions for various initial
conditions.  A finite difference algorithm is used to solve the partial
differential equation for a spherically symmetric system.

The paper is divided in two parts: First, the analytical and numerical
properties of the two flavor Vlasov equation are given. Then this
equation is generalized to include strangeness and also a collision
term is added, so one deals with a Boltzmann type equation with
selfconsistent medium dependent input.

The most serious limitation encountered is the nonrenomalizability of
the NJL model, which is treated by a cutoff of $\Lambda = 602$~MeV. For
a thermal system with $T=240$~MeV about 60\% of all particles have
momenta $p>\Lambda$ and cannot be properly accounted for.

For the Vlasov equation the total energy includes a potential as well
as a kinetic term.  The potential contributes ca.~5\% to the total
energy at $t=0$. During the evolution, its contribution tends to zero
as the mass evolves towards its vacuum value. The expansion dynamics in
the Vlasov equation is dominated by the increase of the constituent
quark mass from $m_q=58$~MeV at $t=0$ to $m_q=312$~MeV at $t=\infty$.
Correspondingly the quarks are slowed down during the expansion. For
this reason the particle spectrum in momentum space changes during the
expansion such that the low momentum region is enhanced and the high
momentum region depleted.

Simultaneously with the introduction of the collision term, the
calculation is generalized to the three flavor model in order to study
also the production of strange quarks. The collision term is
implemented in a relaxation time approximation. In contrast to the
Vlasov equation, the Boltzmann equation leads to an energy spectrum
with an exponential shape at high energies. However, the mean field
still manifests itself in an enhancement of the the low energy region
of the spectra like for the Vlasov equation. The shapes of the spectra
do not change when the medium effects in the cross sections are
switched off.

Only for large systems chemical as well as thermal equilibration of
strange quarks can be reached. The approach to chemical equilibrium as
a function of the initial radius $r_0$ can be well approximated by a
$\tanh(ar_0)$ dependence. The absolute value of the strangeness yield
found in our calculation is about a factor three smaller than observed
in experiment \cite{marek}.

What have we learned from this calculation? Admittedly, an expanding
quark system is an artificial system, since it lacks confinement.  Yet
it is a system in which the dynamics of a non-equilibrium system, which
underlies chiral symmetry, can be studied. Medium effects are taken
care of in the mean field as well as in the collision term. In this
respect the sophistication surpasses other calculations using more
realistic interaction models.  While we cannot compare directly with
experiment, yet at least three results may be of relevance: (i) The
final spectra strongly depend on the mean field. For instance it is
responsible for the enhancement at small energies. This is a direct
consequence of the transition from a chirally symmetric to a chirally
broken phase. (ii) The calculated spectra are rather robust with
respect to changes in the collision terms. Whether one uses medium
dependent cross sections or Born cross sections makes little
difference, provided one does not modify the masses and the mean field.
(iii) Medium effects are small for the masses of the strange quarks.
The rather high values of $m_s$ are one reason, why strangeness
production in the quark plasma alone accounts for only 30\% of the
observed value.

\section*{Acknowledgments}
We wish to thank S.\,P.~Klevansky for illuminating discussions.  This
work has been supported in part by the Deutsche Forschungsgemeinschaft
under contracts no. Hu 233/4-4, and the Federal Ministry for Education
and Research under contract no. 06 HD 742.

\begin{table}
\caption[]{Particle multiplicities for different settings of initial system
           radius and maximal momentum. In row (c), high momenta are cut
           off by a multiplicative factor.}
\label{multitab}
\begin{tabular}{l||c||c|c|c||c|c}
 & $p_{\rm max}/\Lambda$ &
${\cal N}_{q,t=0}$ &
${\cal N}_{q,t=\infty}$ &
${\cal N}_{s,t=\infty}$ &
${\cal N}_{s,t=\infty} / {\cal N}_{q,t=\infty}$ &
$\left({\cal N}_q + {\cal N}_s \right)_{t=\infty} /
{\cal N}_{q,t=0}$ \\ \hline
(a) & 1  & 343.2 & 365.6 &  8.1 & 2.2\% & 1.089 \\
(b) & 3  & 811.2 & 825.4 & 32.4 & 3.9\% & 1.058 \\
(c) & 1* & 251.2 & 242.3 &  5.7 & 2.3\% & 0.987 \\
\end{tabular}
\end{table}

\begin{table}
\caption[]{Dependence of strangeness yield and slope temperatures on the
           medium effects.}
\label{borntab}
\begin{tabular}{l||c|c||c|c|c}
 & Mean Field & Cross Sections &
${\cal N}_{s,t=\infty} / {\cal N}_{q,t=\infty}$ &
$T_q$ (MeV) & $T_s$ (MeV) \\ \hline
(a) & yes & exact & 2.2\% & 185 & 153 \\
(d) & yes & Born  & 3.0\% & 179 & 153 \\
(e) & no  & Born  & 5.5\% & 207 & 191
\end{tabular}
\end{table}

\end{document}